\documentclass[prd,nofootinbib,amsfonts,notitlepage]{revtex4-1}
\usepackage[utf8]{inputenc}
\usepackage[T1]{fontenc}
\usepackage{hyperref}
\usepackage{graphicx,bm,amsmath,color}
\usepackage{bbold}
\usepackage{wasysym}
\usepackage{verbatim}
\usepackage{amssymb}
\usepackage{epstopdf}
\usepackage{animate}
\usepackage{xmpmulti}
\usepackage{centernot}
\usepackage{multirow}
\usepackage{tikz}
\usepackage{graphicx}
\usepackage{float}
\usepackage{mathtools}
\usepackage{comment}
\usetikzlibrary{arrows}
\usetikzlibrary{decorations.pathreplacing}
\DeclareMathOperator{\arccoth}{arccoth}
\DeclareMathOperator{\sech}{sech}
\makeatletter

\newcommand{\be}{\begin{equation}}
\newcommand{\ee}{\end{equation}}
\newcommand{\beq}{\begin{eqnarray}}
\newcommand{\eeq}{\end{eqnarray}}
\newcommand{\ba}{\begin{align}}
\newcommand{\ea}{\end{align}}

\begin{document}

\title{Towards a geometrical interpretation of rainbow geometries\\ for quantum gravity phenomenology}
\author{J.J. Relancio}
\affiliation{Departamento de F\'{\i}sica Te\'orica and Centro de Astropartículas y Física de Altas Energías (CAPA),
Universidad de Zaragoza, Zaragoza 50009, Spain}
\email{relancio@unizar.es}
\author{S. Liberati}
\affiliation{SISSA, Via Bonomea 265, 34136 Trieste, Italy and INFN, Sezione di Trieste;\\ IFPU - Institute for Fundamental Physics of the Universe, Via Beirut 2, 34014 Trieste, Italy}
\email{liberati@sissa.it}

\begin{abstract}
In the literature, there are several papers establishing a correspondence between a deformed kinematics and a nontrivial (momentum dependent) metric. In this work, we study in detail the relationship between the trajectories  given by a deformed Hamiltonian and the geodesic motion  obtained from a geometry in the cotangent bundle, finding that both trajectories coincide when the Hamiltonian is identified with the squared distance in momentum space. Moreover, following the natural structure of the cotangent bundle geometry, we construct the space-time curvature tensor, from which we obtain generalized Einstein equations. Since the metric is not invariant under momentum diffeomorphisms (changes of momentum coordinates) we note that, in order to have a conserved Einstein tensor (in the same sense of general relativity), a privileged momentum basis appears, a completely new result that cannot be found in absence of space-time curvature, which settles a long standing ambiguity of this geometric approach. After that, we consider in an expanding universe the geodesic motion and the Raychaudhuri's equations, and we show how to construct vacuum solutions to the Einstein equations. Finally, we make a comment about the possible phenomenological implications of our framework.  
\end{abstract}

\maketitle

\section{Introduction}
It is well known even to the general public that the main challenge facing theoretical physics in the last decades has been the formulation of a quantum gravity theory (QGT). Indeed, when one attempts to quantize gravity one finds that there are infinities that cannot be canceled as for the rest of interactions~\cite{Feynman:1996kb}. At the root of this problem lies the fact that in quantum field theory (QFT) and general relativity (GR) a very different role is played by spacetime: it is a static frame in QFT and a dynamical variable in GR. 

One of the possibilities considered in many quantum gravity approaches is that the structure of spacetime changes completely as we know it for high energies (or small distances). For example, in loop quantum gravity~\cite{Sahlmann:2010zf,Dupuis:2012yw}, such structure take the form of a spin foam~\cite{Wheeler:1955zz,Rovelli:2002vp,Ng:2011rn,Perez:2012wv}, that can be interpreted as a ``quantum'' spacetime, and in causal set theory~\cite{Wallden:2013kka,Wallden:2010sh,Henson:2006kf} and string theory~\cite{Mukhi:2011zz,Aharony:1999ks,Dienes:1996du}, non-locality effects appears. However, none of these approaches can so far claim to provide a full fledged QGT. Not only we face formidable technical challenges but we are also in the unprecedented condition to attempt to develop a new theory without direct guidance from experiments and observations.

It is partially with the task of ameliorating this situation that in the last two decades a new field of research, broadly labeled quantum gravity phenomenology (QGP), has been developed. In this bottom-up approach, instead of considering a possible fundamental QGT as the ones mentioned above, one attempts to consider systematically all the possible sub-Planck scale manifestations which can be associated to the aforementioned scenarios (where for example the Planck energy is $E_{\rm Pl}=1.2\cdot 10^{28}$ eV). This has the advantage that these approaches may lead to some falsifiable phenomenology, which might serve us as a guidance in building a QGT. 

The basic idea behind this approach is that the transition between a full QGT regime and the classical, smooth, spacetime we do experience daily is not an abrupt transition at the Planck scale but rather a gradual one, which allows for a mesoscopic regime where both the classical and continuous limits (which do not need to coincide) for the spacetime structure and dynamics are not exact. Of course testing these mesoscopic phenomena does not necessary pin points a specific QGT but can strongly constrain (or in case of future detections support) some scenarios and give us some much needed guidance towards theoretical developments.

QGP has in the last two decades focused on several possible phenomenological implications of QG scenarios. More noticeably, the consequences of possible deviations from the local Poincaré invariance of spacetime~\cite{Philpott_2009,Liberati:2013xla,AmelinoCamelia:2008qg}, departures from standard QFT locality (see e.g.~~\cite{Belenchia:2014fda,Belenchia:2015ake}), extra dimensions, or QG modified dynamics in cosmology~\cite{Agullo:2016tjh, Alesci:2016xqa} and in black hole physics~\cite{Rovelli:2014cta,Barcelo:2014cla,Abedi:2016hgu,Carballo-Rubio:2018jzw}.

With regards to the possible deviations from the local Poincaré structure of spacetime, two main scenarios have been intensively studied depending on the fate of the Lorentz symmetry: one can consider that for high energies a Lorentz invariance violation  (LIV)~\cite{Colladay:1998fq,Kostelecky:2008ts,Myers:2003fd} can arise, or that this symmetry is deformed, leading to the theories known as deformed special relativity (DSR)~\cite{AmelinoCamelia:2008qg}. 

The two approaches are not only conceptually alternative but also lead to very distinguished approaches to phenomenology. The preferred frame associated with the breakdown of Lorentz invariance can be characterised within usual Riemannian geometries plus a fundamental vector field~\cite{Jacobson:2000xp, Horava:2009uw}. Deformations of the relativistic group instead cannot be geometrically characterised without introducing a dependence of the spacetime geometry on the energy at which the latter is probed by some specific particle~\cite{Magueijo:2002xx} (a statement that per se is open to different interpretations, e.g.~what determines the relevant energy associated with a given particle, the Compton length or the De Broglie one?). I.e.~ for this class of scenarios one is lead to consider a generalization of the GR pseudo-Riemannian metric that must possess a smooth limit to GR when the velocity or energy of the particle tends to zero.

It is well known that from the space-time manifold one can construct the tangent or cotangent bundle structure~\cite{2012arXiv1203.4101M}. While in GR this is not considered, it is a  crucial point to be taken into account when considering to go beyond this framework with a metric which depends on the velocities or momenta of the particles. This complex structure of tangent or cotangent bundle is not used in GR because a curved spacetime is a particular case of these kind of geometries (where the metric only depends on the space-time coordinates) for which this extra structure would be an unnecessary complication: every result of GR can be obtained considering only the space-time properties. Then, the structure of the tangent and cotangent spaces are in this case trivial due to the absence of velocities and momenta on the metric.

There are several papers in the literature trying to describe a spacetime accounting for these possible generalizations to GR. The possibility of having different velocities for photons depending on their energy was considered through a metric which depends on the vectors of the tangent manifold, i.e. on the velocities. This approach, known as Finsler spaces~\cite{Rund2012}, was studied in several works~\cite{Kostelecky:2011qz,Stavrinos:2016xyg,Hasse:2019zqi}, including some studies of the symmetries in the DSR context~\cite{Girelli:2006fw,Amelino-Camelia:2014rga,Lobo:2016xzq}, making explicit the relativity principle that characterizes these theories.  

Another approach is the one associated to the Hamilton spaces~\cite{2012arXiv1203.4101M}, which is the Hamiltonian version of the Lagrangian formalism followed in the Finsler geometries. In this case, the metric depends, besides the space-time coordinates, on the cotangent vectors, i.e. on the momentum of the particle probing the spacetime. This kind of geometry was considered in the DSR context in~\cite{Barcaroli:2015xda,Barcaroli:2016yrl}. 

In both Finsler and Hamilton geometries, the main ingredient from which a nontrivial metric results, depending on the velocity or momentum respectively, is a deformed dispersion relation. Then, if one considers in both cases the so-called ``classical basis'' of $\kappa$-Poincaré~\cite{Borowiec2010}, in which the dispersion relation is the usual one of SR, no velocity or momentum dependence of the metric appears to arise. Moreover, as discussed in the introduction, the main ingredient of DSR is a deformation of the momentum conservation laws, and in the aforementioned approaches it is not clear how to account for such a feature.

In~\cite{Carmona:2019fwf} it was rigorously shown that a de Sitter momentum space leads to $\kappa$-Poincaré kinematics identifying the isometries and the squared distance of the metric with the main ingredients of the kinematics: translations, Lorentz isometries and the square of the distance can be interpreted as the deformed composition law, deformed Lorentz transformations and deformed dispersion relation respectively (the last two facts where previously contemplated in Refs.~\cite{AmelinoCamelia:2011bm,Lobo:2016blj}). In that proposal, all the kinematical ingredients that characterize DSR are accounted for. Moreover, when one considers the ``classical basis'' of $\kappa$-Poincaré in this approach, the momentum metric is still nontrivial. 

While  from both geometrical and algebraical point of view different choices of the kinematics of $\kappa$-Poincaré (different choices of coordinates in a de Sitter momentum space or different bases in Hopf algebras~\cite{KowalskiGlikman:2002we}) represent the same deformed kinematics (with the same properties, such as the associativity of the composition law and the relativity principle), there is an ambiguity about what are the momentum variables associated to physical measurements. The fact that different bases could represent different physics was deeply considered in the literature~\cite{AmelinoCamelia:2010pd}. In this work, we propose a way to select this ``physical'' basis from geometrical considerations.

The work developed in~\cite{Carmona:2019fwf} was generalized in~\cite{Relancio:2020zok} in order to include a curvature in spacetime. There, the present authors realized that the metric formalism considered in~\cite{Carmona:2019fwf} can be seen as a particular case of a cotangent bundle geometry~\cite{2012arXiv1203.4101M}, which is a generalization of a Hamilton space. Moreover we found that, with a particular prescription of this metric in the cotangent bundle, one finds that the same trajectories of particles are obtained from the line element and from the Casimir defined to be the squared distance in momentum space. In this work we study in more detail this relationship.

The structure of the paper is as follows. In Sec.~\ref{sec:hamilton-metric}, after explaining the main ingredients of a cotangent bundle geometry that we will use along the paper, we show the relationship between the Hamilton equations with a deformed Casimir and the geodesic motion obtained in this geometrical framework. In Sec.~\ref{sec:different_considerations} we compute the Lie derivative and how to include an external force in this formalism. At the end of the section we review how to construct a metric in the cotangent bundle we used in~\cite{Relancio:2020zok}, and how this is related with the results of the previous section. We define the covariant derivative along a curve in Sec.~\ref{sec:curvature_tensors}. This allows us to define a Riemann tensor (which can be also obtained from the commutator of two space-time covariant derivatives) and then, an Einstein tensor. Since in our proposal there is an invariance under space-time diffeomorphisms but not under change of momentum coordinates, the fact that the Einstein tensor is conserved (in order to be also conserved the energy momentum tensor) selects one and only one momentum coordinates describing a de Sitter momentum space. We use this basis in order to compute the modified Raychaudhuri’s equations in Sec.~\ref{sec:raychaudhuri} and study the geodesics in an universe expanding in Sec.~\ref{sec:universe}. In  Sec.~\ref{sec:vacuum} show how to consider  any vacuum solution of Einstein equations, including a cosmological constant term. Finally, we end with the conclusions, discussing the possible phenomenology of our proposal.

\section{Relationship between Hamilton equations and geodesics from a cotangent bundle metric}
\label{sec:hamilton-metric}
Let us start by reviewing the main elements of a geometry in the cotangent bundle and showing that there is a clear correspondence between the equations of motion obtained from a Hamiltonian and the trajectory of particles given by a geodesic equation.

\subsection{Main properties of the geometry in the cotangent bundle}

In reviewing the main properties of the cotangent bundle structure, we can start from the case of GR so to make the reader familiar with the approach we will follow along the paper.

We start by considering a base manifold $M$ with $n$  dimensions. In the GR case, the base manifold is the curved spacetime in which particles move and $n=4$. From this manifold, one can construct the cotangent bundle manifold $T^*M$, which has dimension $2n$. This manifold is formed by the base manifold (the spacetime) and the fibers (the momentum space). Then, this structure gives a geometry for all the phase-space variables. As we have said in the introduction, the fiber bundle is not usually considered in GR. However, it is necessary for considering a momentum dependent metric.

\begin{figure}[h]
\centering
\begin{tikzpicture}
\draw (-4,1.5) node (v2) {} -- (-5.5,-0.5) node (v1) {};
\draw  plot[smooth, tension=.7] coordinates {(v1) (-1.5,-0.5) (2,-2)};
\draw  plot[smooth, tension=.7] coordinates {(v2)};
\draw  plot[smooth, tension=.7] coordinates {(v2) (-0.5,1.5) (3,0)};
\draw (2,-2) -- (3,0);
\draw  plot[smooth, tension=.7] coordinates {(-2.5,1.6) (-3.3,0.6) (-4,-0.4)};
\draw  plot[smooth, tension=.7] coordinates {(-1.1,1.6) (-1.75,0.65) (-2.4,-0.4)};
\draw  plot[smooth, tension=.7] coordinates {(0.4,1.2) (-0.15,0.25) (-0.7,-.73)};
\draw  plot[smooth, tension=.7] coordinates {(1.75,0.68) (1.2,-0.35) (0.7,-1.35)};
\draw  plot[smooth, tension=.7] coordinates {(-4.7,0.6) (-1,0.5) (2.5,-1)};
\node (v4) at (-1,0.36) {};
\node (v3) at (-1,3.1) {};
\node (v5) at (-1.55,2.15) {$\partial/\partial k_\nu$};
\draw [-latex](v4) -- (v3);
\node (v6) at (4,-0.5) {};
\node (v8) at (3.3,-0.75) {$\delta/\delta x^\mu$};
\node (v9) at (-1.15,0.51) {};
\draw [-latex](v9) -- (v6);
\node (v7) at (4,0.5) {};
\draw [-latex](v9) -- (v7);
\node at (3.3,0.8) {$\partial/\partial x^\mu$};
\draw [decorate,decoration={brace,amplitude=5pt,mirror,raise=0pt},yshift=0pt]
(4,-0.5) -- (4,0.5) node [font=\bfseries,midway,xshift=1.1cm,yshift=1pt]
{$N_{\mu \nu}\partial/\partial k_\nu$};
\node at (-2,-1) {$M$};
\node at (-6.8,1.4) {$V$};
\draw  plot[smooth, tension=.7] coordinates {(v1) (-6,1.5) (-5.5076,3)};
\draw  plot[smooth, tension=.7] coordinates {(-4.7,0.6) (-5.1,2.05) (-4.5,3.5)};
\draw  plot[smooth, tension=.7] coordinates {(v2) (-4.25,2.55) (-3.5,4)};
\node[draw,align=left] at (-0.35,4.2) {$T^*M$};
\end{tikzpicture}
\caption{Visualization of the cotangent structure.}
\label{figure}
\end{figure}
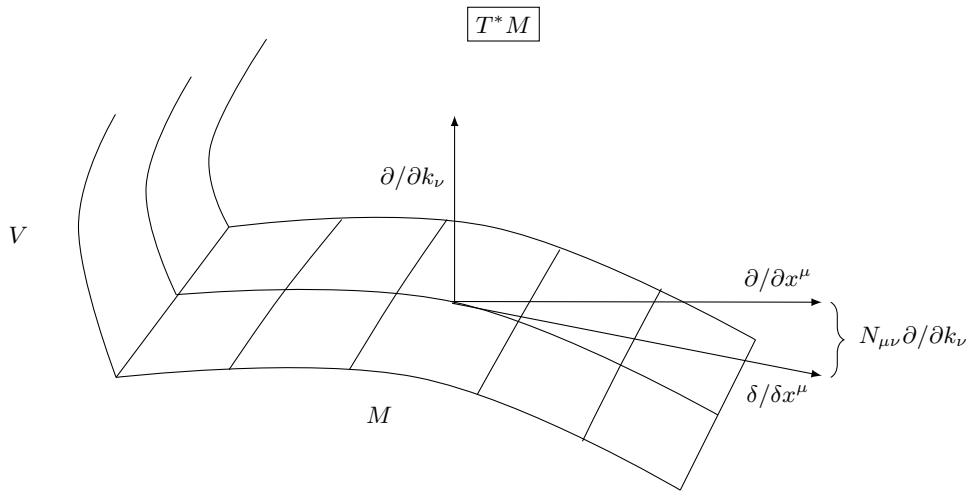

On the cotangent bundle manifold, associating to each point $u \in T^*M$ (i.e. a point in phase space $(x,k)$) the fiber $V_u$ (all the points with fixed $x$ but different $k$), one can obtain the so called vertical distribution, $V: u \in T^*M\rightarrow V_u \subset T_u T^*M$ with dimension $n$, which is generated by $\partial/\partial k$. As it is shown in Fig~\ref{figure}, given a point on the cotangent bundle one can construct the vertical distribution (the fiber). Note that in the figure the fiber is unidimensional for the sake of simplicity, but in fact it has the same dimensions as the base manifold.

As it is shown in~\cite{2012arXiv1203.4101M} one can define a  nonlinear connection $N$ (also called horizontal distribution), supplementary to the vertical distribution $V$, i.e. $ T_u T^*M=N_u\oplus V_u$, which also has dimension $n$. One can construct an adapted basis for the horizontal distribution
\begin{equation}
\frac{\delta}{\delta x^\mu}\, \doteq \,\frac{\partial}{\partial x^\mu}+N_{\nu\mu}(x,k)\frac{ \partial}{\partial k_\nu}\,,
\label{eq:delta_derivative}
\end{equation}  
where $N_{ \nu \mu}$ are the coefficients of the nonlinear connection. The choice of these coefficients is not unique but, as it is shown in~\cite{2012arXiv1203.4101M}, there is one and only one choice of nonlinear connection coefficients that leads to a space-time affine connection which is metric compatible and torsion free.
In GR, the coefficients of the nonlinear connection are given by
\begin{equation}
N_{\mu\nu}(x,k)\, = \, k_\rho \Gamma^\rho_{\mu\nu}(x)\,,
\label{eq:nonlinear_connection}
\end{equation} 
where $ \Gamma^\rho_{\mu\nu}(x)$ is the affine connection. Then, when the metric does not depend on the space-time coordinates these coefficients vanish. 

We can visualize the difference between the horizontal distribution generated by $\delta/\delta x$ and the curves generated by $\partial/\partial x$ in Fig~\ref{figure}. The horizontal distribution is chosen to be supplementary to the vertical distribution, making that moving along the vector $\delta/\delta x$ suppose a movement not only in the base manifold (the spacetime) but in along the fiber (momentum space). We can regard the case of GR as an example: along a geodesic the momentum of the particle changes, provoking a movement along the fiber. While this movement is trivial in GR since the metric does not depend on the momentum of the particle, this is not the case for a metric which depends on all phase-space coordinates, as we will see in the following. 

In~\cite{2012arXiv1203.4101M} a line element in the cotangent bundle is defined as  
\begin{equation}
\mathcal{G}\,=\, g_{\mu\nu}(x,k) dx^\mu dx^\nu+g^{\mu\nu}(x,k) \delta k_\mu \delta k_\nu\,,
\label{eq:line_element_ps} 
\end{equation}
where 
\begin{equation}
\delta k_\mu \,=\, d k_\mu - N_{\nu\mu}(x,k)\,dx^\nu\,. 
\end{equation}

The \textit{d-curvature tensor} is defined as~\cite{2012arXiv1203.4101M}
\begin{equation}
R_{\mu\nu\rho}(x,k)\,=\,\frac{\delta N_{\nu\mu}(x,k)}{\delta x^\rho}-\frac{\delta N_{\rho\mu}(x,k)}{\delta x^\nu}\,.
\label{eq:dtensor}
\end{equation} 
It represents the curvature of the phase space. It can be seen that this tensor in GR is proportional to the Riemann tensor 
\begin{equation}
R_{\mu\nu\rho}(x,k)\,=\,k_\sigma R^{\sigma}_{\mu\nu\rho}(x)\,.
\end{equation} 
It measures the integrability of spacetime, i.e. position space, as a subspace of the cotangent bundle and is defined as the commutator between the horizontal vector fields
\begin{equation}
\left[ \frac{\delta}{\delta x^\mu}\,,\frac{\delta}{\delta x^\nu}\right]\,=\,\frac{\delta}{\delta x^\mu}\frac{\delta}{\delta x^\nu}-\frac{\delta}{\delta x^\nu}\frac{\delta}{\delta x^\mu}\,=\, R_{\mu\nu\rho}(x,k)\frac{\partial}{\partial k_\rho}\,.
\label{eq:commutator_deltas}
\end{equation}

In~\cite{2012arXiv1203.4101M} it is shown that a vertical path is characterized as a curve in the cotangent bundle with constant space-time coordinates and with the momentum satisfying the geodesic equation with the connection of the momentum space, i.e.
\begin{equation}
x^\mu\left(\tau\right)\,=\,x^\mu_0\,,\qquad \frac{d^2k_\mu}{d\tau^2}+C_\mu^{\nu\sigma}(x_0,k)\frac{dk_\nu}{d\tau}\frac{dk_\sigma}{d\tau}\,=\,0\,,
\end{equation} 
where 
\begin{equation}
C_\rho^{\mu\nu}(x,k)\,=\,-\frac{1}{2}g_{\rho\sigma}\left(\frac{\partial g^{\sigma\nu}(x,k)}{\partial k_ \mu}+\frac{\partial g^{\sigma\mu}(x,k)}{\partial k_ \nu}-\frac{\partial g^{\mu \nu}(x,k)}{\partial k_ \sigma}\right)\,,
\label{eq:affine_connection_p}
\end{equation}
is affine connection in momentum space, while a horizontal curve will be determined by the geodesic motion in spacetime given by
\begin{equation}
\frac{d^2x^\mu}{d\tau^2}+H^\mu_{\nu\sigma}(x,k)\frac{dx^\nu}{d\tau}\frac{dx^\sigma}{d\tau}\,=\,0\,,
\label{eq:horizontal_geodesics_curve_definition}
\end{equation} 
where 
\begin{equation}
H^\rho_{\mu\nu}(x,k)\,=\,\frac{1}{2}g^{\rho\sigma}(x,k)\left(\frac{\delta g_{\sigma\nu}(x,k)}{\delta x^\mu} +\frac{\delta g_{\sigma\mu}(x,k)}{\delta  x^\nu} -\frac{\delta g_{\mu\nu}(x,k)}{\delta x^\sigma} \right)\,,
\label{eq:affine_connection_st}
\end{equation} 
is the affine connection of spacetime
and the change of momentum obtained from
\begin{equation}
\frac{\delta k_\lambda}{d \tau}\,=\,\frac{dk_\lambda}{d\tau}-N_{\sigma\lambda} (x,k)\frac{dx^\sigma}{d\tau}\,=\,0\,.
\label{eq:horizontal_momenta}
\end{equation} 
Here, $\tau$ plays the role of the proper time or the affine parametrization depending if one is considering a massive or a massless particle respectively. 

In~\cite{2012arXiv1203.4101M} it was defined the covariant derivatives in space-time 
\begin{equation}
\begin{split}
T^{\alpha_1 \ldots\alpha_r}_{\beta_1\ldots\beta_s;\mu}(x,k)\,&=\,\frac{\delta T^{\alpha_1 \ldots\alpha_r}_{\beta_1\ldots\beta_s}(x,k)}{\delta x^\mu}+T^{\lambda \alpha_2 \ldots\alpha_r}_{\beta_1\ldots\beta_s}(x,k)H^{\alpha_1}_{\lambda \mu}(x,k)+\cdots+T^{\alpha_1 \ldots \lambda}_{\beta_1\ldots\beta_s}(x,k)H^{\alpha_r}_{\lambda \mu}(x,k)\\
&-T^{\alpha_1 \ldots \alpha_r}_{\lambda \beta_2\ldots\beta_s}(x,k)H^{\lambda}_{\beta_1 \mu}(x,k)-\cdots-T^{\alpha_1 \ldots \alpha_r}_{\beta_1\ldots \lambda}(x,k)H^{\lambda}_{\beta_s \mu}(x,k)\,,
\label{eq:cov_dev_st}
\end{split}
\end{equation} 
and in momentum space
\begin{equation}
\begin{split}
T^{\alpha_1 \ldots\alpha_r;\mu}_{\beta_1\ldots\beta_s}(x,k)\,&=\,\frac{\partial T^{\alpha_1 \ldots\alpha_r}_{\beta_1\ldots\beta_s}(x,k)}{\partial k_\mu}+T^{\lambda \alpha_2 \ldots\alpha_r}_{\beta_1\ldots\beta_s}(x,k)C^{\alpha_1\mu}_{\lambda}(x,k)+\cdots+T^{\alpha_1 \ldots \lambda}_{\beta_1\ldots\beta_s}(x,k)C^{\alpha_r \mu}_{\lambda }(x,k)\\
&-T^{\alpha_1 \ldots \alpha_r}_{\lambda \beta_2\ldots\beta_s}(x,k)C^{\lambda \mu}_{\beta_1 }(x,k)-\cdots-T^{\alpha_1 \ldots \alpha_r}_{\beta_1\ldots \lambda}(x,k)C^{\lambda \mu}_{\beta_s}(x,k)\,.
\label{eq:cov_dev_ms}
\end{split}
\end{equation} 
Also, it is shown that given a metric, there is always a symmetric nonlinear connection leading to the affine connections in space-time and in momentum spaces making that both covariant derivatives of the metric vanishes:
\begin{equation}
g_{\mu\nu;\rho}\,=\,g_{\mu\nu}^{\,\,\,\,\,\,;\rho}\,=\,0\,.
\label{eq:covariant_derivative_2}
\end{equation} 

\subsection{Geodesics from Hamilton equations}
\label{sub:geo}
We shall now show that the above given geodesic equation can be recovered, consistently, from the Hamilton equations as well as from the action.  We start by computing the equations of motion from a Hamiltonian $\mathcal{C}$ (which would be the dispersion relation $\mathcal{C} (x,k)=f(x,k,m,\Lambda)$):
\begin{equation}
\frac{dx^\mu}{d\tau}\,=\,\mathcal{N}\lbrace{\mathcal{C} (x,k),x^\mu\rbrace}\,=\,\mathcal{N}\frac{\partial \mathcal{C} (x,k)}{\partial k_\mu}\,,\qquad\frac{dk_\mu}{d\tau}\,=\,\mathcal{N}\lbrace{\mathcal{C} (x,k),k_\mu\rbrace}\,=\,-\mathcal{N}\frac{\partial \mathcal{C} (x,k)}{\partial x^\mu}\,,
\label{eq:H-eqs}
\end{equation}
where $\mathcal{N}$ is a Lagrange multiplier, being $1/2m$ or $1$ for massive and massless particles respectively and we used the Poisson bracket 
\begin{equation}
\lbrace{f,g\rbrace}\,=\,\frac{\partial f}{\partial k_\rho}\frac{\partial g}{\partial x^\rho}-\frac{\partial g}{\partial k_\rho}\frac{\partial f}{\partial x^\rho}\,.
\label{eq:pb}
\end{equation}

We know that for a horizontal curve Eq.~\eqref{eq:horizontal_momenta} holds, so substituting Eq.~\eqref{eq:H-eqs} in Eq.~\eqref{eq:horizontal_momenta} we obtain 
\begin{equation}
 \frac{\delta k_\lambda}{d \tau}\,=\,\frac{dk_\lambda}{d\tau}-N_{\sigma\lambda} (x,k)\frac{dx^\sigma}{d\tau}\,=\,-\mathcal{N}\left(\frac{\partial \mathcal{C} (x,k)}{\partial x^\lambda}+N_{\sigma\lambda}\frac{\partial \mathcal{C} (x,k)}{\partial k_\sigma} \right)\,=\,0\,,
\label{eq:derivative_casimir}
 \end{equation}
which implies
\begin{equation}
 \frac{\delta \mathcal{C} (x,k)}{\delta x^\mu}\,=\,\mathcal{C} _{;\mu}(x,k)\,=\,0\,.
\label{eq:delta_casimir}
\end{equation}
This result was obtained from different considerations in~\cite{Barcaroli:2015xda}. This is consistent with the fact that the Casimir is a constant along geodesics and then, its covariant derivative should vanish. Note that this is not the case when considering the usual covariant derivative of GR (i.e. the partial with respect to the coordinates). In fact, it is easy to check that Eq.~\eqref{eq:delta_casimir} is satisfied in GR for the Casimir $\mathcal{C} (x,k)=k_\mu g^{\mu\nu}(x)k_\nu$:
\begin{equation}
\begin{split}
&k_\lambda \frac{\partial g^{\lambda\sigma}(x)}{\partial x^\mu}k_\sigma+2k_\lambda \Gamma^\lambda_{\mu \rho}(x)g^{\rho \sigma}(x)k_\sigma\,=\,k_\lambda k_\sigma\left(\frac{\partial g^{\lambda\sigma}(x)}{\partial x^\mu}+g^{\lambda \tau}(x)g^{\rho \sigma}(x)\left(\frac{\partial g_{\mu \tau}(x)}{\partial x^\rho}+\frac{\partial g_{\rho\tau}(x)}{\partial x^\mu}-\frac{\partial g_{\mu\rho}(x)}{\partial x^\tau}\right)\right)\,=\,\\
&k_\lambda k_\sigma\left(\frac{\partial g^{\lambda\sigma}(x)}{\partial x^\mu}-g_{\mu \tau}(x)g^{\rho \sigma}(x)\frac{\partial g^{\lambda \tau}(x)}{\partial x^\rho}-g_{\rho\tau}(x)g^{\rho \sigma}(x)\frac{\partial g^{\lambda \tau}(x)}{\partial x^\mu}+g^{\lambda \tau}(x)g_{\mu\rho}(x)\frac{\partial g^{\rho \sigma}(x)}{\partial x^\tau}\right)\,=\,0\,,
\end{split}
\label{eq:delta_casimir_GR}
\end{equation}
where we have used Eq.~\eqref{eq:nonlinear_connection}, the explicit form of the affine connection in GR as a function of the metric and the symmetry under the exchange of $\lambda \leftrightarrow \sigma$. 

Now we can compute the second derivative of the position finding
\begin{equation}
\frac{d^2x^\mu}{d\tau^2}\,=\, \mathcal{N}\frac{d}{d\tau}\frac{\partial \mathcal{C} (x,k)}{\partial k_\mu}\,=\, \mathcal{N}\left(\frac{\partial^2 \mathcal{C} (x,k)}{\partial k_\mu \partial x^\rho} \frac{dx^\rho}{d\tau}+\frac{\partial^2 \mathcal{C} (x,k)}{\partial k_\mu \partial k_\sigma} \frac{dk_\sigma}{d\tau}\right)\,.
\label{eq:geo_1}
\end{equation}
The first term in the right-hand side of the previous equation can be rewritten using Eq.~\eqref{eq:derivative_casimir} as 
\begin{equation}
\frac{\partial^2 \mathcal{C} (x,k)}{\partial k_\mu \partial x^\rho}\,=\, \frac{\partial }{\partial k_\mu} \frac{\partial \mathcal{C} (x,k)}{\partial x^\rho}\,=\, - \frac{\partial }{\partial k_\mu} N_{\rho \sigma}(x,k) \frac{\partial \mathcal{C} (x,k)}{\partial k_\sigma}\,=\,- \frac{\partial N_{\rho \sigma}(x,k)}{\partial k_\mu}  \frac{\partial \mathcal{C} (x,k)}{\partial k_\sigma}- N_{\rho \sigma}(x,k) \frac{\partial^2 \mathcal{C} (x,k)}{ \partial k_\mu \partial k_\sigma}\,.
\end{equation}

Then, using the relation~\eqref{eq:horizontal_momenta}, Eq.~\eqref{eq:geo_1} becomes
\begin{equation}
\frac{d^2x^\mu}{d\tau^2}\,=\,  \mathcal{N}\left(- \frac{\partial N_{\rho \sigma}(x,k)}{\partial k_\mu}  \frac{\partial \mathcal{C} (x,k)}{\partial k_\sigma}- N_{\rho \sigma}(x,k) \frac{\partial^2 \mathcal{C} (x,k)}{\partial k_\mu\partial k_\sigma}+\frac{\partial^2 \mathcal{C} (x,k)}{\partial k_\mu \partial k_\sigma} N_{\rho \sigma} (x,k)\right) \frac{dx^\rho}{d\tau}\,=\,-  \mathcal{N} \frac{\partial N_{\rho \sigma}(x,k)}{\partial k_\mu}  \frac{\partial \mathcal{C} (x,k)}{\partial k_\sigma} \frac{dx^\rho}{d\tau}\,.
\end{equation}

Finally, using the first expression of Eq.\eqref{eq:H-eqs} one can easily find that 
\begin{equation}
\frac{d^2x^\mu}{d\tau^2}+H^\mu_{\nu\sigma}(x,k)\frac{dx^\nu}{d\tau}\frac{dx^\sigma}{d\tau}\,=\,0\,,
\label{eq:horizontal_geodesics}
\end{equation}
where 
\begin{equation}
H^\mu_{\nu\sigma}(x,k)\,=\,\frac{\partial N_{\nu\sigma}(x,k)}{\partial k_\mu}\,.
\label{eq:affine_connection_n}
\end{equation}
When the Casimir is a quadratic expression of the momentum, one obtains that $H^\mu_{\nu\sigma}(x,k)$ is the usual affine connection of GR. In that case, one finds Eq.~\eqref{eq:nonlinear_connection}

At the beginning of the previous subsection we mentioned that there is one and only one  nonlinear connection that leads to a space-time affine connection which is metric compatible, making that the covariant derivative of the metric vanishes~\eqref{eq:covariant_derivative_2}.  While in Hamilton spaces there is a clear formula relating the nonlinear coefficients with the metric and the Hamiltonian~\cite{2012arXiv1203.4101M}, this is not the case for a generalized Hamilton space. In~\cite{2012arXiv1203.4101M} it was pointed out for this case one cannot determine the nonlinear connection from the metric. However, whilst  in this approach  there is not a single formula from which we can obtain these coefficients, using Eqs.~\eqref{eq:affine_connection_st},\eqref{eq:delta_casimir},\eqref{eq:affine_connection_n}, one can determine them.

However, until now we have not established a connection between the metric and the Casimir. Since our starting point in our previous works~\cite{Relancio:2020zok,Relancio:2020mpa} was a momentum dependent metric, we need to relate the Casimir with the metric in order to determine the space-time affine connection and the nonlinear connection directly from the geometry. This is the goal of the next subsection.

\subsection{Conservation of the Casimir along an horizontal path}
\label{sec:casimir_horizontal}

Here we show that the distance in momentum space is conserved along a horizontal path. Then, one can consider a function of such distance as the Casimir. 

For a horizontal curve, we know that an infinitesimal displacement leads to
\begin{equation}
x^{\prime \mu}\,=\,x^\mu + \Delta x^\mu\,,\qquad k^{\prime }_\nu\,=\,k_\nu + N_{\mu \nu} (x,k)\,\Delta x^\mu\,.
\label{eq:variation_horizontal}
\end{equation}
If the momentum line elements of a phase-space point $(x,k)$ and another infinitesimal near to this one along a horizontal curve $(x^\prime,k^\prime)$ are the same (i.e. the distances in momentum space from $(x,0)$ to $(x,k)$ and  $(x^\prime,0)$ to  $(x^\prime,k^\prime)$ are the same), the following equation must hold
\begin{equation}
dk^\prime_\mu g^{\mu \nu} (x^\prime,k^\prime)dk^\prime_\nu\,=\,dk_\rho g^{\rho \sigma} (x,k) dk_\sigma\,.
\label{eq:momentum_isometry}
\end{equation}
We can rewrite the previous expression as a function of the metric with subindexes finding
\begin{equation}
g_{\mu \nu} (x^\prime,k^\prime)\,=\, \frac{\partial k^\prime_\mu }{\partial k_\rho} g_{\rho \sigma} (x,k)\frac{\partial k^\prime_\nu }{\partial k_\sigma} \,,
\label{eq:momentum_isometry2}
\end{equation}
and using Eq.~\eqref{eq:variation_horizontal} we finally obtain
\begin{equation}
\frac{\delta g_{\mu \nu}(x,k)}{\delta x^\rho}-\frac{\partial N_{\mu \rho}(x,k)}{\partial k_\lambda} g_{\lambda \nu}(x,k)-\frac{\partial N_{\nu \rho}(x,k)}{\partial k_\lambda} g_{\lambda \mu}(x,k)\,=\,\frac{\delta g_{\mu \nu}(x,k)}{\delta x^\rho}-H^\lambda_{\mu \rho}(x,k) g_{\lambda \nu}(x,k)-H^\lambda_{\nu \rho}(x,k) g_{\lambda \mu}(x,k)\,=\,0\,,
\label{eq:covariant_derivative}
\end{equation}
where in the second equation we have used the relation~\eqref{eq:affine_connection_n} between the space-time affine connection and the nonlinear connection and in the last step that the space-time covariant derivative of the metric vanishes~\eqref{eq:covariant_derivative_2}.

Although here we have shown that the Casimir can be identified with any function of the distance, in Sec.\ref{sec:construction_metric} we shall see that, with the proposal of how to construct the metric in the cotangent bundle used in~\cite{Relancio:2020zok}, the Casimir has to be the squared of the distance in momentum space. This will allow us to define unequivocally the space-time affine connection and the nonlinear connection from our particular construction of a metric in the cotangent bundle.

\subsection{Geodesics from action}

In this subsection, instead of starting by a Casimir, let us consider the same action of GR regarding the line element, that is we check if in our formalism geodesics are still straightforward and shortest curves as in metric theories of gravity
\begin{equation}
S\,=\,\int ds\,=\,\int \sqrt{dx^\mu g_{\mu \nu}(x,k)dx^\nu}\,,
\label{eq:action}
\end{equation}
where now the metric depends also on the momentum. We know from the definition of the phase-space line element~\eqref{eq:line_element_ps} that, since we want to study horizontal curves, the derivative of the momentum satisfies Eq.~\eqref{eq:horizontal_momenta}. Making a variation $\Delta x^\rho$, we find
\begin{equation}
 \Delta S\,=\,\int \Delta \sqrt{\frac{dx^\mu}{d \tau} g_{\mu \nu}(x,k)\frac{dx^\nu}{d \tau} } d\tau\,=\,0.
\label{eq:action2}
\end{equation}
This implies that 
\begin{equation}
\int \left(\frac{dx^\mu}{d \tau} \Delta g_{\mu \nu}(x,k)\frac{dx^\nu}{d \tau}+2 \frac{d \Delta x^\mu}{d \tau} g_{\mu \nu}(x,k)\frac{dx^\nu}{d \tau}\right) d\tau\,=\,0\,.
\label{eq:variation1}
\end{equation}

For the first term we take into account that 
\begin{equation}
\Delta g_{\mu \nu}(x,k)\,=\,\frac{\partial g_{\mu \nu}(x,k)}{\partial x^\rho} \Delta x^\rho+\frac{\partial g_{\mu \nu}(x,k)}{\partial k_\sigma} \Delta k_\sigma\,,
\label{eq:variation_metric}
\end{equation}
where, using Eq.~\eqref{eq:horizontal_momenta}
\begin{equation}
\Delta k_\sigma\,=\,N_{\rho \sigma} (x,k) \Delta x^\rho\,,
\end{equation}
and then, we can rewrite Eq.~\eqref{eq:variation_metric} as 
 \begin{equation}
\Delta g_{\mu \nu}(x,k)\,=\,\frac{\delta  g_{\mu \nu}(x,k)}{\delta x^\rho} \Delta x^\rho\,.
\label{eq:variation_metric2}
\end{equation}

We can now integrate by parts the second term in~\eqref{eq:variation1} obtaining 
\begin{equation}
\int \left(2 \frac{d \Delta x^\mu}{d \tau} g_{\mu \nu}(x,k)\frac{dx^\nu}{d \tau}\right) d\tau\,=\,-2\int\Delta x^\rho \left(\frac{\delta g_{\rho \nu}(x,k)}{\delta x^\sigma} \frac{d x^\sigma}{d\tau}\frac{d x^\nu}{d\tau}+g_{\rho \nu}(x,k)\frac{d^2 x^\nu}{d\tau^2}\right) d\tau \,.
\end{equation}

Combining both terms we find that, in order to be the variation of the action null, the term proportional to $\Delta x^\rho$ must vanish since the result cannot depend on the variation. Then, we have  
\begin{equation}
2\frac{\delta g_{\rho \nu}(x,k)}{\delta x^\sigma} \frac{d x^\sigma}{d\tau}\frac{d x^\nu}{d\tau}+2g_{\rho\nu}(x,k)\frac{d^2 x^\nu}{d\tau^2}-\frac{\delta  g_{\mu \nu}(x,k)}{\delta x^\rho}\frac{dx^\mu}{d \tau}\frac{dx^\nu}{d \tau}\,=\,0\,.
\end{equation} 
Multiplying the previous equation by $1/2\, g^{\mu \rho}(x,k)$ and rearranging indexes we obtain
\begin{equation}
\frac{d^2 x^\mu}{d\tau^2}+\frac{1}{2} g^{\mu \rho}(x,k) \left(2\frac{\delta g_{\rho \nu}(x,k)}{\delta x^\sigma}-\frac{\delta  g_{\sigma \nu}(x,k)}{\delta x^\rho}\right)\frac{dx^\nu}{d \tau}\frac{dx^\sigma}{d \tau}\,=\,0\,.
\end{equation} 
Due to the symmetry of the previous equation by interchanging the indexes $\nu$ and $\sigma $, we can decompose the first term in the parenthesis into two terms finally finding 
\begin{equation}
\frac{d^2x^\mu}{d\tau^2}+H^\mu_{\nu\sigma}(x,k)\frac{dx^\nu}{d\tau}\frac{dx^\sigma}{d\tau}\,=\,0\,,
\label{eq:geodesic_equation}
\end{equation} 
where $H^\mu_{\nu\sigma}(x,k)$ is the same expression~\eqref{eq:affine_connection_st} obtained in~\cite{2012arXiv1203.4101M}.

Also, starting from an action for the vertical curves (moving along the fiber for a fixed space-time point)
\begin{equation}
S\,=\,\int ds\,=\,\int \sqrt{dk_\mu g^{\mu \nu}(x,k)dk_\nu}\,,
\label{eq:action_momentum}
\end{equation}
one can find that the momentum affine connection is the one of Eq.~\eqref{eq:affine_connection_p}.

\section{Different considerations on the metric formalism}
\label{sec:different_considerations}
In this section, we find the Lie derivative of a vector in this context and how to include an external force in our formalism. We also review the particular construction of the metric in the cotangent bundle that we considered in~\cite{Relancio:2020zok} and that we shall use in the following, explaining its main properties and characteristics.

\subsection{Lie derivative}
In this subsection we derive the modified Lie derivative for a contravariant vector in the cotangent space since we will use this result in Sec.~\ref{sec:raychaudhuri}. We can express the variation of the coordinates $x^\alpha$ along a vector field $\xi^\alpha(x)$ as 
\begin{equation}
x^{\prime \alpha}(x)\,=\,x^\alpha+\xi^\alpha(x) \Delta\lambda\,,
\end{equation}
where $\lambda$ is he infinitesimal variation parameter. This variation of $x^\alpha$ reflects on $k_\alpha$ in the following way
\begin{equation}
k^\prime_{\alpha}\,=\,k_\beta \frac{\partial x^\beta}{\partial x^{\prime \alpha}}\,=\,k_\alpha-\frac{\partial\xi^\beta(x)}{\partial x^\alpha}k_\beta \Delta\lambda\,,
\label{eq:x_variation}
\end{equation}
since $k$ transforms as a covector. The general variation of a vector field $u^\alpha\left(x,k\right)$ will then be
\begin{equation}
\Delta u^\alpha(x,k)\,=\,\frac{\partial u^\alpha(x,k)}{\partial x^\beta} \Delta x^\beta+\frac{\partial u^\alpha(x,k)}{\partial k_\beta} \Delta k_\beta\,=\,\frac{\partial u^\alpha(x,k)}{\partial x^\beta}\xi^\beta (x) \Delta\lambda-\frac{\partial u^\alpha(x,k)}{\partial k_\beta}\frac{\partial\xi^\gamma (x) }{\partial x^\beta}k_\gamma \Delta\lambda \,.
\label{eq:vector_variation}
\end{equation}

Then, the Lie derivative is defined as 
\begin{equation}
\mathcal{L}_\xi u^\mu(x,k)\,=\, \frac{u^{\prime \mu}(x^\prime,k^\prime)-u^\mu(x^\prime,k^\prime)}{\Delta \lambda}\,=\,\xi^\nu(x) \frac{\partial u^\mu (x,k)}{\partial x^\nu}-u^\nu (x,k)\frac{\partial \xi^\mu (x)}{\partial x^\nu}-\frac{\partial u^\mu (x,k)}{\partial k_\alpha}\frac{\partial \xi^\gamma(x)}{\partial x^\alpha} k_\gamma\,.
\label{eq:lie_def}
\end{equation}
Using the definition~\eqref{eq:cov_dev_st} for the covariant derivative and taking into account that the vector field $\xi(x)$ does not depend on the momentum, one can finally write 
\begin{equation}
\mathcal{L}_\xi u^\mu(x,k)\,=\,\xi^\nu(x) u^\mu_{\,;\nu}(x,k)-u^\nu(x,k) \xi^\mu_{\,;\nu}(x)-\frac{\partial u^\mu(x,k)}{\partial k_\alpha} \frac{\partial \xi^\gamma(x)}{\partial x^\alpha} k_\gamma -\frac{\partial u^\mu(x,k)}{\partial k_\alpha}N_{\alpha \gamma}(x,k)\xi^\gamma(x)\,.
\label{eq:lie_vec}
\end{equation}

It is easy to see that in the limit of a $\Lambda$ independent vector field $u^\mu(x)$ the usual expression for the Lie derivative is recovered. 

\subsection{External forces and fields}
We  can now explore how the geodesic equation is modified when an external force (or field) is acting on it. We have seen that for a horizontal curve there is a relationship between the Casimir and the nonlinear connection in Sec.~\ref{sub:geo}. From that relation we have found that the usual geodesic equation holds (Eq.~\eqref{eq:horizontal_geodesics}) but with a space-time affine connection depending also on the momenta. Also, we can see that the nonlinear connection defines this affine connection in a clear and simple way (Eq.~\eqref{eq:affine_connection_n}). All of the above relations have the usual GR limit for a standard (quadratic in momenta) Casimir. Then, when including an external force, the new terms cannot be included in the nonlinear connection (as proposed in~\cite{2012arXiv1203.4101M}) because in the GR limit one would not obtain the usual geodesic equation in presence of an external force. We can then conclude that the nonlinear connection must be the same one obtained in absence of such force.

Given these premises, in order to find the equations of motion in presence of an external force we shall start by considering the action
\begin{equation}
S\,=\,\int \left(m \sqrt{\frac{dx^\mu}{d \tau} g_{\mu \nu}(x,k)\frac{dx^\nu}{d \tau}}-\mathcal{L}\left(x,dx/d \tau\right)\right)d \tau\,,
\label{eq:action_external}
\end{equation}
where the last term represent the Lagrangian associated to the external force. This action is defined as a generalization of the usual one in the presence of external forces. From the variation of this action one can find 
\begin{equation}
\frac{d^2x^\mu}{d\tau^2}+H^\mu_{\nu\sigma}(x,k)\frac{dx^\nu}{d\tau}\frac{dx^\sigma}{d\tau}\,=\,g^{\mu \nu}(x,k)f_\nu\left(x,dx/d \tau\right)\,,
\label{eq:horizontal_geodesics_external}
\end{equation}
being 
\begin{equation}
f_\nu\left(x,dx/d \tau\right)\,=\, \frac{\partial \mathcal{L}\left(x,dx/d \tau\right)}{\partial x^\nu}-\frac{d}{d\tau}\left(\frac{\partial \mathcal{L}\left(x,dx/d \tau\right)}{\partial \left( dx^\nu/d\tau \right)}\right)
\end{equation}
the acceleration induced by the external force.

In this way, we have generalized the usual GR result to the case in which the space-time affine connection and the metric depend on the momentum of the particle.

To be concrete, we can now introduce an electromagnetic field depicted by a four-vector $A_\mu (x)$. In this case, Eq.~\eqref{eq:action_external} becomes
\begin{equation}
S\,=\,\int \left(m \sqrt{\frac{dx^\mu}{d \tau} g_{\mu \nu}(x,k)\frac{dx^\nu}{d \tau}}- q\frac{dx^\mu}{d \tau}A_\mu (x)\right)d \tau\,.
\label{eq:action_em}
\end{equation}
It is easy to find that the trajectories of particles are then 
\begin{equation}
\frac{d^2x^\mu}{d\tau^2}+H^\mu_{\nu\sigma}(x,k)\frac{dx^\nu}{d\tau}\frac{dx^\sigma}{d\tau}\,=\,\frac{q}{m} g^{\mu \lambda}(x,k)F_{\lambda \nu}(x) \frac{dx^\nu}{d\tau}\,,
\label{eq:horizontal_geodesics_em}
\end{equation}
where
\begin{equation}
F_{\lambda \nu}(x)\,=\,\frac{\partial A_\nu(x)}{\partial x^\lambda}-\frac{\partial A_\lambda(x)}{\partial x^\nu}
\end{equation}
is the usual electromagnetic tensor.

These are the same equations used in~\cite{Relancio:2020mpa} in order to describe the motion of a charged particle with a momentum dependent metric. 

\subsection{Construction of the metric in the cotangent bundle}
\label{sec:construction_metric}
In~\cite{Relancio:2020zok} we proposed that the metric tensor $g_{\mu\nu}(x,k)$ has to be constructed in a very particular way with the tetrads of spacetime and momentum space. Explicitly,
\begin{equation}
g_{\mu\nu}(x,k)\,=\,\Phi^\alpha_\mu(x,k) \eta_{\alpha\beta}\Phi^\beta_\nu(x,k)\,,
\label{eq:cotangent_metric_tetrads}
\end{equation}
where 
\begin{equation}
\Phi^\alpha_\mu(x,k)\,=\,e^\lambda_\mu(x)\varphi^\alpha_\lambda(\bar{k})\,,
\label{eq:tetrad_cotangent}
\end{equation}
being $e^\lambda_\mu(x)$ the tetrad of spacetime, $\varphi^\alpha_\lambda(k)$ the tetrad in momentum space (which we choose to represent a maximally symmetric momentum space for reasons we see in the point 2 of the following enumeration) and $\bar{k}_\mu=\bar{e}^\lambda_\mu(x) k_\lambda$, where $\bar{e}^\lambda_\mu(x)$ is the inverse of the space-time tetrad.  

This particular choice of the metric has some interesting properties:
\begin{enumerate}
\item It is easy to see that the line element of Eq.~\eqref{eq:line_element_ps} is invariant when using the metric of Eq.~\eqref{eq:cotangent_metric_tetrads}. Indeed~\cite{Relancio:2020zok} 
\begin{equation}
\Phi'^\mu_\rho(x',k')\,=\, \frac{\partial x^\nu}{\partial x'^\rho} \Phi^\mu_\nu(x,k)\,,
\label{eq:tetrad_canonical_transformation}
\end{equation}
because
\begin{equation}
\frac{\partial x^\mu}{\partial x'^\rho}  e^\lambda_\mu(x)\varphi^\alpha_\lambda(\bar{k})\,=\,e'^\kappa_\rho(x')\varphi'^\alpha_\kappa(\bar{k}')\,,
\label{eq:tetrad_transformation_fragmented}
\end{equation}
where we used a standard transformation law for the tetrad of spacetime
\begin{equation}
\bar{e}'^\nu_\mu(x')\,=\,\frac{\partial x'^\nu}{\partial x^\rho}\bar{e}^\rho_\mu(x)\,.
\end{equation} 
Then, the barred momentum variables are independent of the choice of spatial coordinates 
\begin{equation}
\bar{k}'_ \mu\,=\,k'_\nu \bar{e}'^\nu_\mu(x')\,=\,\frac{\partial x^\sigma}{\partial x'^\nu}k_\sigma \frac{\partial x'^\nu}{\partial x^\rho} \bar{e}'^\rho_\mu(x')\,=\,k_\nu \bar{e}^\nu_\mu(x) \,=\,\bar{k}_ \mu\,.
\end{equation} 
Also, we have considered that the momentum space tetrad do not change under such transformation. 

\item In~\cite{Carmona:2019fwf} it was shown that the most known deformed kinematics ($\kappa$-Poincaré~\cite{Majid1994}, Snyder~\cite{Battisti:2010sr} and hybrid models~\cite{Meljanac:2009ej}) can be described within this framework by considering the geometrical properties of a maximally symmetric momentum space. This can be achieved by defining a deformed composition and transformation laws from the isometries of the momentum metric associated to translations and Lorentz respectively. Similarly, a deformed dispersion relation can be seen as the (square of the) distance from the origin to a point in momentum space. Explicitly, the composition law satisfies
\begin{equation}
g_{\mu\nu}\left(p\oplus q\right) \,=\,\frac{\partial \left(p\oplus q\right)_\mu}{\partial q_\rho} g_{\rho\sigma}(q)\frac{\partial \left(p\oplus q\right)_\nu}{\partial q_\sigma}\,,
\label{eq:composition_isometry}
\end{equation} 
which, in order to be associative~\cite{Carmona:2019fwf}, requires the following relation between the (inverse of the) tetrad in momentum space and the composition law to hold
\begin{equation}
\varphi^\mu_\nu(p \oplus q) \,=\,  \frac{\partial (p \oplus q)_\nu}{\partial q_\rho} \, \varphi_\rho^{\,\mu}(q)\,.
\label{eq:tetrad_composition}
\end{equation}
The modified Lorentz transformations are instead given  by the six isometries leaving invariant the origin
\begin{equation}
\frac{\partial g^k_{\mu\nu}(k)}{\partial k_\rho} {\cal J}^{\alpha\beta}_\rho(k) \,=\,
\frac{\partial{\cal J}^{\alpha\beta}_\mu(k)}{\partial k_\rho} g^k_{\rho\nu}(k) +
\frac{\partial{\cal J}^{\alpha\beta}_\nu(k)}{\partial k_\rho} g^k_{\mu\rho}(k)\,,
\label{eq:cal(J)}
\end{equation}
where
\begin{equation}
 {\cal J}^{\alpha\beta}(k) \,=\,x^\mu {\cal J}^{\alpha\beta}_\mu(k)\,,
\end{equation}
is the Lorentz generator~\cite{Carmona:2019fwf}.

Our proposal for constructing a metric in the whole cotangent bundle, preserves that the momentum curvature tensor also gives a constant scalar of curvature~\cite{Relancio:2020zok}, so allowing to introduce 10 momentum isometries leading to the same kinematics obtained in~\cite{Carmona:2019fwf} in the flat space-time limit~\cite{Relancio:2020zok}. In particular, we saw for the particular case in which the composition law is associative that, defining $p\rightarrow \bar{p}_\mu=\bar{e}_\mu^\nu(x_0) p_\nu$, $q\rightarrow \bar{q}_\mu=\bar{e}_\mu^\nu(x_0) q_\nu$ for a fixed space-time point $x_0$, the modified composition ($\bar{\oplus}$)
\be
(\bar{p} \oplus \bar{q})_\mu \,=\, \bar{e}_\mu^\nu(x_0) (p \bar{\oplus} q)_\nu\,,
\label{eq:composition_cotangent}
\ee
is an isometry of the tetrad~\eqref{eq:tetrad_cotangent}
\begin{equation}
\Phi^\mu_\nu(x_0,(p \bar{\oplus} q)) \,=\,  \frac{\partial (p \bar{\oplus} q)_\nu}{\partial q_\rho} \, \Phi_\rho^{\,\mu}(x_0,q)\,,
\label{eq:tetrad_cotangent_composition}
\end{equation}
and then, an isometry of the metric in the cotangent bundle (for a fixed space-time point).

\item Also, we obtained in~\cite{Relancio:2020zok} a simple relationship between the metric and the Casimir defined as the distance squared when the metric does not depend on the space-time coordinates
 \begin{equation}
\frac{\partial C(k)}{\partial k_ \mu}g_{\mu\nu}(k) \frac{\partial C(k)}{\partial k_ \nu}\,=\,4 C(k)\,.
\label{eq:casimir_definition}
\end{equation}
With this fact we showed that considering an action with a deformed dispersion relation
and a momentum geometry where one identifies the squared distance with the Casimir, leads to the same results.

Moreover, with a metric constructed as in Eq.~\eqref{eq:cotangent_metric_tetrads} we found that the same equation holds 
 \begin{equation}
\frac{\partial C(\bar{k})}{\partial k_ \mu}g_{\mu\nu}(x,k) \frac{\partial C(\bar{k})}{\partial k_\nu}\,=\,4 C(\bar{k})\,,
\label{eq:casimir_definition_cst}
\end{equation}
where the new Casimir is now a function of the barred momenta, being this the squared of the distance in momentum space for a fixed space-time point for a metric constructed as in Eq.~\eqref{eq:cotangent_metric_tetrads}. We observed the same relationship between the action and metric formalisms even when considering a nontrivial geometry for the spacetime~\cite{Relancio:2020zok}. Then, compatibility of our proposal for constructing the metric in the cotangent bundle with the result obtained in Sec.~\ref{sec:casimir_horizontal}, requires the Casimir  to be the squared of the distance in momentum space.  Note that when considering a momentum independent metric, one finds the usual results of SR and GR for flat and curved spacetime respectively.

\end{enumerate}

\section{Space-time and momentum curvature tensors}
\label{sec:curvature_tensors}
In this section we discuss how to obtain the space-time curvature tensor when considering a nontrivial (momentum dependent) metric in the cotangent bundle. As we will see, to impose that the contracted Bianchi identity holds and that the Einstein tensor is conserved selects one and only one momentum basis. This momentum basis would be the one in which every study and phenomenological computation should be carried out (even for a flat spacetime, in order to have a smooth limit). 
\label{sec:riemann}
\subsection{Covariant derivative along a curve}
A possible way to derive the modified Riemann tensor in spacetime is through the geodesic deviation. However, in order to study this we need first to generalize the concept of directional derivative. We first start by considering that, as in GR~\cite{Weinberg:1972kfs}, a tensor $A^\mu(\tau)$ transforms under a diffeomorphism as 
\begin{equation}
A^{\prime \mu}(\tau)\,=\, \frac{\partial x^{\prime \mu}}{\partial x^\nu}A^\nu(\tau)\,,
\end{equation}
so
\begin{equation}
\frac{d A^{\prime \mu}(\tau)}{d\tau}\,=\, \frac{\partial x^{\prime \mu}}{\partial x^\nu}\frac{d A^\nu(\tau)}{d\tau}+\frac{\partial^2 x^{\prime \mu}}{\partial x^\nu\partial x^\rho}\frac{dx^\rho}{d\tau}A^\nu(\tau)\,.
\end{equation}

Since the modified affine connection~\eqref{eq:affine_connection_st} transforms as the usual connection of GR~\cite{2012arXiv1203.4101M}, it is then natural to define a covariant derivative along a curve $x^\mu(\tau)$ as 
\begin{equation}
\frac{D A^{\mu}(\tau)}{D\tau}\,\coloneqq\,\frac{d A^{\mu}(\tau)}{d\tau}+H^\mu_{\nu \rho}(x,k)\frac{dx^\rho}{d\tau}A^\nu(\tau) \,.
\label{cov_der_curve}
\end{equation}
In~\cite{Rund2012} it was also used an analogous expression for a Finsler space. Note that this allows us to write the geodesic equation as in GR
 \begin{equation}
\frac{D u^{\mu}}{D\tau} \,=\,\frac{\partial u^\mu }{\partial x^\nu}\frac{d x^\nu}{d \tau}+\frac{\partial u^\mu }{\partial k_\rho }\frac{d k_\rho}{d \tau}+H^{\mu}_{\nu \sigma}(x,k) u^\nu u^\sigma \,=\,\frac{\delta u^\mu }{\delta x^\nu} u^\nu+H^{\mu}_{\nu \sigma}(x,k) u^\nu u^\sigma \,=\,u^\mu_{;\nu}u^\nu\,=\,0\,, 
\label{eq:geodesic_cov_curve}
\end{equation}
being $u^\mu=dx^\mu/d\tau$ and where we have used that for a horizontal curve~\eqref{eq:horizontal_momenta} holds.

\subsection{Geodesic deviation}
We can now try to obtain the space-time Riemann tensor from the geodesic deviation. Let us consider a pair of nearby freely falling particles with trajectories $x^\mu(\tau)$ and $x^\mu(\tau)+\Delta x^\mu(\tau)$. The equations of motion are
 \begin{equation}
\frac{d^2x^\mu}{d\tau^2}+H^\mu_{\nu\sigma}(x,k)\frac{dx^\nu}{d\tau}\frac{dx^\sigma}{d\tau}\,=\,0\,,\qquad\frac{d^2 \left(x^\mu+\Delta x^\mu\right)}{d\tau^2}+H^\mu_{\nu\sigma}(x+\Delta x,k+\Delta k)\frac{d\left(x^\nu+\Delta x^\nu\right)}{d\tau}\frac{d\left(x^\sigma+\Delta x^\sigma\right)}{d\tau}\,=\,0\,, 
\end{equation}
where $\Delta k$ is given by Eq.~\eqref{eq:variation_horizontal} since we are considering a horizontal curve. Then, we can compute the difference between both trajectories at first order in $\Delta x^\mu$ obtaining 
 \begin{equation}
\frac{d^2 \left(x^\mu+\Delta x^\mu\right)}{d\tau^2}+\frac{\delta H^\mu_{\nu\sigma}(x,k)}{\delta x^\rho}\Delta x^\rho\frac{dx^\nu}{d\tau}\frac{dx^\sigma}{d\tau}+ H^\mu_{\nu\sigma}(x,k)\frac{dx^\nu}{d\tau}\left( \frac{dx^\sigma}{d\tau}+2 \frac{d\Delta x^\sigma}{d\tau}\right)\,=\,0\,.
\end{equation}
Using the geodesic equation~\eqref{eq:horizontal_geodesics_curve_definition}, we can simplify the previous expression, getting
 \begin{equation}
\frac{d^2 \Delta x^\mu}{d\tau^2}+\frac{\delta H^\mu_{\nu\sigma}(x,k)}{\delta x^\rho}\Delta x^\rho\frac{dx^\nu}{d\tau}\frac{dx^\sigma}{d\tau}+ 2 H^\mu_{\nu\sigma}(x,k)\frac{dx^\nu}{d\tau} \frac{d\Delta x^\sigma}{d\tau}\,=\,0\,.
\label{eq:inter_cov}
\end{equation}
Now, from the definition of covariant derivative along a curve~\eqref{cov_der_curve}, we can compute
 \begin{equation}
\begin{split}
\frac{D^2 \Delta x^\mu}{D\tau^2}\,=\,&\frac{d}{d\tau}\left(\frac{d \Delta x^\mu}{d \tau}+H^\mu_{\nu\rho}(x,k)\frac{d x^\rho}{d \tau}\Delta x^\nu\right)+H^\mu_{\nu\rho}(x,k)\frac{d x^\nu}{d \tau}\left(\frac{d \Delta x^\rho}{d \tau}+H^\rho_{\sigma\lambda}(x,k)\frac{d x^\sigma}{d \tau}\Delta x^\lambda\right)\,=\\ \nonumber
& \frac{d^2 \Delta x^\mu}{d\tau^2}+\frac{\delta H^\mu_{\nu\rho}}{\delta x^\sigma}\frac{d x^\sigma}{d\tau}\frac{d x^\rho}{d\tau}\Delta x^\nu +2 H^\mu_{\nu\rho}\frac{d x^\nu}{d\tau}\frac{d \Delta x^\rho}{d\tau}+ H^\mu_{\sigma\rho} H^\rho_{\nu\lambda}\Delta x^\nu \frac{d x^\lambda}{d\tau}\frac{d x^\sigma}{d\tau}+H^\mu_{\nu\rho} \Delta x^\nu\ \frac{d^2  x^\nu}{d\tau^2}
\,.
\end{split}
\end{equation}

Finally, using Eqs.~\eqref{eq:horizontal_geodesics_curve_definition}-\eqref{eq:inter_cov} we can obtain the Riemann tensor from 
 \begin{equation}
\frac{D^2 \Delta x^\mu}{D\tau^2}\,=\,R^\mu_{\nu \rho\sigma}(x,k)\Delta x^\rho \frac{dx^\nu}{d\tau}\frac{dx^\sigma}{d\tau}\,,
\label{eq:deo_dev_eq}
\end{equation}
where 
 \begin{equation}
R^\mu_{\nu \rho\sigma}(x,k)\,=\,\frac{\delta H^\mu_{\nu \rho}(x,k)}{\delta x^\sigma}-\frac{\delta H^\mu_{\nu \sigma}(x,k)}{\delta x^\rho}+H^\lambda_{\nu \rho}(x,k)H^\mu_{\lambda \sigma}(x,k)-H^\lambda_{\nu \sigma}(x,k)H^\mu_{\lambda \rho}(x,k)\,.
\label{riemann_st}
\end{equation}
Note that this definition of the Riemann tensor differs from the one defined in~\cite{2012arXiv1203.4101M}. When computing the commutator of two space-time covariant derivatives of a vector field $X^\mu (x,k)$ one obtains (when there is no torsion)
 \begin{equation}
X^\mu (x,k)_{;\nu;\rho}-X^\mu (x,k)_{;\rho;\nu}\,=\,X^\lambda (x,k) R^{*\mu}_{\lambda \nu\rho}(x,k)-X^\mu (x,k)^{;\lambda} R_{\lambda \nu\rho}(x,k) \,, 
\label{commutator_cov_der2}
\end{equation}
where $R_{\lambda \nu\rho}(x,k)$ is the \textit{d-curvature tensor} tensor defined in Eq.~\eqref{eq:dtensor} and
 \begin{equation}
R^{*\mu}_{\lambda \nu\rho}(x,k)\,=\,R^{\mu}_{\lambda \nu\rho}(x,k)+C^{\mu \sigma}_\lambda (x,k)R_{\sigma \nu\rho}(x,k)\,.
\end{equation}
This last tensor is what in~\cite{2012arXiv1203.4101M} is defined as the Riemann tensor. However, note that this definition is ad hoc since one can rewrite Eq.~\eqref{commutator_cov_der2} as well as
 \begin{equation}
X^\mu (x,k)_{;\nu;\rho}-X^\mu (x,k)_{;\rho;\nu}\,=\,X^\lambda (x,k) R^{\mu}_{\lambda \nu\rho}(x,k)-\frac{\partial X^\mu (x,k)}{\partial p_\lambda} R_{\lambda \nu\rho}(x,k) \,. 
\label{commutator_cov_der}
\end{equation}
Our definition of the Riemann tensor from the geodesic deviation corresponds to identify the Riemann tensor with the term proportional to the field in the difference of two covariant derivatives as normally done in (pseudo-) Riemannian geometries.

\subsection{Momentum curvature tensor}
One can obtain also the curvature tensor in momentum space by computing the commutator of two momentum covariant derivatives~\cite{2012arXiv1203.4101M}
 \begin{equation}
X^{\mu;\nu;\rho} (x,k)-X^{\mu;\rho;\nu}(x,k)\,=\,X^\lambda (x,k) S^{\mu\nu\rho}_{\lambda }(x,k) \,, 
\label{commutator_cov_der2_momentum}
\end{equation}
where
\begin{equation}
S_{\sigma}^{\mu\nu\rho}(x,k)\,=\, \frac{\partial C^{\mu\nu}_\sigma(x,k)}{\partial k_\rho}-\frac{\partial C^{\mu\rho}_\sigma(x,k)}{\partial k_\nu}+C_\sigma^{\lambda\nu}(x,k)C^{\mu\rho}_\lambda(x,k)-C_\sigma^{\lambda\rho}(x,k)C^{\mu\nu}_\lambda(x,k)\,.
\label{eq:Riemann_p}
\end{equation} 
This is the result obtained in~\cite{2012arXiv1203.4101M} when the affine connection in momentum space is symmetric by the exchange of its upper indexes. 
\section{About the Einstein equations}
\label{sec:einstein}
In this section we find a privileged basis thanks to a modification of the Einstein's equations and we discuss the possible choice of momentum basis and the invariance of the model under momentum diffeomorphisms.

\subsection{Privileged basis from Einstein's equations}
We can now define an Einstein tensor (as it was proposed in~\cite{2012arXiv1203.4101M}) from the Riemann tensor 
 \begin{equation}
G_{\mu \nu}(x,k)\,=\,R_{\mu \nu}(x,k)-\frac{1}{2}g_{\mu \nu}(x,k)R(x,k) \,, 
\label{eq:einstein_tensor}
\end{equation}
where
 \begin{equation}
R_{\mu \nu}(x,k)\,=\, R^{\lambda}_{\mu \nu\lambda }(x,k)\,,\qquad R(x,k)\,=\,R_{\mu \nu} (x,k)g^{\mu \nu}(x,k)\,.
\label{eq:ricci_scalar}
\end{equation}
Following the physical idea that the momentum dependence of the metric is some sort of mesoscopic regime which should smoothly recover a Riemannian geometry at low energies, it is natural to impose that the Einstein equations continue to hold. So that
 \begin{equation}
R_{\mu \nu}(x,k)-\frac{1}{2}g_{\mu \nu}(x,k)R(x,k) \,=\,8\pi T_{\mu \nu}(x,k)\,.
\label{eq:einstein_eqs}
\end{equation}
Requiring that the energy-momentum tensor is conserved
 \begin{equation}
T^\mu _{\nu;\mu}(x,k)\,=\,0\,, 
\label{eq:conservation_em_tensor}
\end{equation}
implies by consistency that the second Bianchi identity holds
 \begin{equation}
R^\lambda_{ \mu \nu \kappa;\eta } (x,k)+R^\lambda_{\mu \eta\nu ; \kappa} (x,k)+R^\lambda_{\mu \kappa\eta ; \nu} (x,k)\,=\,0\,.
\label{eq:second_bianchi}
\end{equation}
While this equation is automatically satisfied in GR, this is no more granted in this context. Indeed, for~\eqref{eq:second_bianchi} to hold, one finds that the following combination must vanish
\begin{equation}
R_{\mu \lambda\tau} (x,k)\frac{\partial H^\nu_{\rho\sigma}(x,k)}{\partial k_\mu}+R_{\mu \tau \rho} (x,k)\frac{\partial H^\nu_{\lambda\sigma}(x,k)}{\partial k_\mu}+R_{\mu \rho\lambda} (x,k)\frac{\partial H^\nu_{\sigma\tau}(x,k)}{\partial k_\mu}\,=\,0\,.
\label{eq:second_bianchi_condition}
\end{equation}
These terms appear since, while the commutator of two space-time partial derivatives is null, this is not the case for the commutator of two $\delta$ derivatives, as it is shown in Eq.~\eqref{eq:commutator_deltas}. 

Eq.~\eqref{eq:second_bianchi_condition} is obviously satisfied if $H^\nu_{\rho\sigma}(x,k)$ does not depend on $k$. Furthermore, when one imposes that the Einstein tensor is conserved, i.e.
\begin{equation}
\left(R^{\mu \nu}(x,k)-\frac{1}{2}g^{\mu \nu}(x,k)R(x,k)\right)_{;\mu}\,=\,0\,,
\label{eq:cov_der_Einstein}
\end{equation}
one finds that only one basis satisfies this property, i.e. the one in which the momentum metric (i.e. in absence of space-time curvature) becomes conformally flat
\begin{equation}
g_{\mu \nu}(k)\,=\,\eta_{\mu\nu}\left(1-\frac{k_0^2-\vec{k}^2}{4\Lambda^2}\right)^2\,.
\label{eq:physical_metric}
\end{equation}
Moreover we know that tensors, and in particular the Einstein one, are not invariant under momentum diffeomorphisms and then, if the conservation of the Einstein tensor holds for one momentum representation, it would not be the case for any other momentum basis. Then, for any other curvature of spacetime this would be also the preferred basis since it is the only one in which the Einstein tensor will be conserved. 

For this metric, one can see that the scalar of curvature in momentum space is constant $S=12/\Lambda^2$ and that indeed the curvature tensor in momentum space corresponds to a maximally symmetric space, i.e.
\begin{equation}
S_{\rho\sigma\mu\nu}\,\propto \, g_{\rho\mu}g_{\sigma\nu}-g_{\rho\nu}g_{\sigma\mu}\,.
\end{equation}
Then, from this metric one can define a relativistic deformed kinematics following the same prescription formulated in~\cite{Carmona:2019fwf}. 

This particular form of the metric implies that the Casimir is a function of $k^2$. The explicit expression can be obtained from Eq.~\eqref{eq:casimir_definition}
\begin{equation}
\mathcal{C}(k)\,=\,4 \Lambda^2 \arccoth^2\left(\frac{2 \Lambda}{\sqrt{k_0^2-\vec{k}^2}}\right)\,.
\label{eq:casimir_physical}
\end{equation}

The associative composition law (corresponding to the coproduct of the momenta of $\kappa$-Poincaré in this basis) can be obtained from Eq.~\eqref{eq:composition_isometry} order by order. In particular, up to second order, the composition law reads
\begin{equation}
\begin{split}
\left(p\oplus q\right)_0\,&=\,p_0+q_0+\frac{\vec{p}\cdot \vec{q}}{\Lambda}-\frac{p_0 q_0 \left(p_0+q_0\right)}{4\Lambda^2}+\frac{\vec{p}\cdot \vec{q} \left(q_0-p_0\right)}{2\Lambda^2}+\frac{3\vec{p}^2 q_0}{4\Lambda^2}-\frac{\vec{q}^2 p_0}{4\Lambda^2}\,,\\
\left(p\oplus q\right)_i\,&=\,p_i+q_i+\frac{p_i q_0}{\Lambda}+p_i\left(\frac{q_0^2}{4\Lambda^2}-\frac{p_0 q_0}{2\Lambda^2}+\frac{\vec{p}\cdot \vec{q}}{2\Lambda^2}-\frac{\vec{q}^2}{4\Lambda^2}\right)+q_i\left(\frac{\vec{p}^2}{4\Lambda^2}-\frac{p_0^2}{4\Lambda^2}-\frac{p_0 q_0}{2\Lambda^2}+\frac{\vec{p}\cdot \vec{q}}{2\Lambda^2}\right)\,.
\end{split}
\end{equation}

The modification of the metric~\eqref{eq:physical_metric} resides on the mass of the particle probing the spacetime. This is consistent with the relativity principle holding in DSR: the only invariant (for the one-particle system) under change of reference systems is the mass of the particle. 

This momentum basis choice implies that the Einstein tensor does not depend on $k$. Indeed, on one hand, the Riemann and then the Ricci tensors, do not depend on $k$, since the space-time affine connection is only a function of the space-time coordinates. On the other hand, due to the particular form of the metric, the term proportional to the product of the metric and the Ricci scalar also turns out to be momentum independent given that, 
\begin{equation}
g_{\mu \nu}(x,k) R(x,k)\,=\,g_{\mu \nu}(x,k)g^{\rho \sigma}(x,k) R_{\rho \sigma}(x)\,=\, \left(1-\frac{\bar{k}_0^2-\vec{\bar{k}}^2}{4\Lambda^2}\right)^2 g^x_{\mu \nu}(x) \left(1-\frac{\bar{k}_0^2-\vec{\bar{k}}^2}{4\Lambda^2}\right)^{-2} g^{\rho \sigma}_x(x)  R_{\rho \sigma}(x)\,=\,g^x_{\mu \nu}(x) R^x(x)\,,
\end{equation}
due to the particular form of the metric~\eqref{eq:physical_metric} 
\begin{equation}
g_{\mu \nu}(x,k)\,=\,\left(1-\frac{\bar{k}_0^2-\vec{\bar{k}}^2}{4\Lambda^2}\right)^2 g^x_{\mu \nu}(x) \,,
\label{eq:metric_cotangent_definition}
\end{equation}
and where $g^x_{\mu \nu}(x)$ and $ R^x(x)$ are the metric and scalar curvature in GR. Then, the Einstein's equations in this scheme take the usual expression of GR. Moreover, in this particular case, since the space-time affine connection does not depend on the momentum, being the usual connection of GR, the nonlinear coefficients are obtained as in GR by Eq.~\eqref{eq:nonlinear_connection}, which satisfies Eq.~\eqref{eq:affine_connection_n}. We can now wonder whether these coefficients satisfy the condition~\eqref{eq:delta_casimir}. We can see from Eq.~\eqref{eq:delta_casimir}
\begin{equation}
 0\,=\,\frac{\delta \mathcal{C} (x,k)}{\delta x^\mu}\,=\,\mathcal{N}\left( \frac{\partial \mathcal{C} (x,k)}{\partial x^\mu}+N_{\mu\nu}(x,k)\frac{\partial \mathcal{C} (x,k)}{\partial k_\nu}\right)\,=\,\mathcal{N}\frac{\partial \mathcal{C} (x,k)}{\partial \bar{k}_\rho}\left(\frac{\partial \bar{k}_\rho}{\partial x^\mu}+N_{\mu\nu}(x,k)\frac{\partial \bar{k}_\rho}{\partial k_\nu}\right)\,,
\label{eq:casimir_delta_2}
\end{equation}
where we have used the fact that the Casimir depends on the phase-space coordinates through $\bar{k}$. For the particular case we are considering, the Casimir~\eqref{eq:casimir_physical} is a function of $k^2$. Then, the generalization of this Casimir for any curved spacetime is the same function of $\bar{k}^2$ (as we have explained in Sec.~\ref{sec:construction_metric}). We can then rewrite Eq.~\eqref{eq:casimir_delta_2} as 
\begin{equation}
\frac{\partial \mathcal{C} (x,k)}{\partial \bar{k}^2}\eta^{\rho \sigma}\bar{k}_\sigma\left(\frac{\partial \bar{k}_\rho}{\partial x^\mu}+N_{\mu\nu}(x,k)\frac{\partial \bar{k}_\rho}{\partial k_\nu}\right)\,=\,\frac{\partial \mathcal{C} (x,k)}{\partial \bar{k}^2}\left(\frac{1}{2}k_\lambda \frac{\partial g^{\lambda\sigma}(x)}{\partial x^\mu}k_\sigma+ N_{\mu \rho}(x,k)g^{\rho \sigma}(x)k_\sigma\right)\,=\,0 \,,
\end{equation}
since the term in parenthesis vanishes when the nonlinear connection is defined by Eq.~\eqref{eq:nonlinear_connection}, as we saw in Eq.~\eqref{eq:delta_casimir_GR}.  Then, when the Casimir is a function of $\bar{k}^2$, the condition~\eqref{eq:delta_casimir} is automatically satisfied for the GR nonlinear connection~\eqref{eq:nonlinear_connection}. With this we have shown that the nonlinear connection~\eqref{eq:nonlinear_connection} is compatible with Eqs.~\eqref{eq:delta_casimir}-\eqref{eq:affine_connection_n}, and then with the metric in the cotangent space~\eqref{eq:metric_cotangent_definition}.

\subsection{About the choice of momentum basis}
In DSR (in flat spacetime) there is a controversy about the different choice of momentum bases and its implications about the phenomenology. In fact, there is an open debate in the literature regarding the possibility that different bases could represent different physics~\cite{AmelinoCamelia:2010pd}. 

Regarding our approach, whilst we have shown in Sec.~\ref{sec:construction_metric} that the line element~\eqref{eq:line_element_ps} is invariant under space-time diffeomorphisms but not under momentum change of variables, it is important to note that in the limit case in which the metric in phase space does not depend on the space-time coordinates, an invariance under different choices of momentum basis arises. This can be easily shown  considering  a canonical transformation in phase space $(x, k) \to (x', k')$ 
\be
k^\prime_\mu \,=\, f_\mu(k)\,,\quad x^{\prime\mu}\,=\, x^\nu h^\mu_\nu(k)\,,
\ee
for any non linear change of momentum variables, i.e, for any set of functions $f_\mu$ of the momentum variables, with
\be
h^\mu_\rho(k)  \,=\, \frac{\partial k_\rho}{\partial  k^\prime_\mu}\,.
\ee
Under this kind of transformation, the line element in phase space is invariant
\begin{equation}
\begin{split}
&g^\prime_{\mu\nu}(k^\prime) dx^{\prime\mu} dx^{\prime\nu}+g^{\prime \mu\nu}(k^\prime) d k^\prime_\mu d k^\prime_\nu\,=\,\frac{\partial x^{\prime\mu}}{\partial x^\rho} g^\prime_{\mu\nu}(k^\prime) \frac{\partial x^{\prime\nu}}{\partial x^\sigma} dx^{\rho} dx^{\sigma}+\frac{\partial k^\prime_\mu}{\partial k_\rho} g^{\prime \mu\nu}(k^\prime)\frac{\partial k^\prime_\nu}{\partial k_\sigma} d k_\rho d k_\sigma\,=\, \\
&g_{\mu\nu}(k) dx^{\mu} dx^{\nu}+g^{ \mu\nu}(k) d k_\mu d k_\nu\,.
\end{split}
\label{eq:line_element_ps_flat} 
\end{equation}

Nevertheless, this cannot be done for the line element~\eqref{eq:line_element_ps}. Then, our model presents a degeneracy when the metric depends only on momentum, making that any momentum coordinates can be used. But when including a curvature on spacetime, one and only one momentum basis is compatible with the conservation of Einstein's equations. Note that this degeneracy appears also when the momentum metric is flat, making that the usual variables of SR (in which the conservation law for momenta is the sum) are the only ones in which the Einstein's equations are conserved.

\section{Congruence of geodesics and Raychaudhuri's equation}
\label{sec:raychaudhuri}
Let us now consider the Raychaudhuri's equation for a metric in the cotangent space. For the sake of simplicity we shall not write explicitly the space-time and momentum dependence of the tensors. We start by computing the expansion of both timelike and null geodesics by its definition from the metric~\cite{Poisson:2009pwt} (as we did in~\cite{Relancio:2020zok}). The expansion of timelike geodesics is 
\begin{equation}
\theta \,=\,\frac{1}{\delta V}\frac{d }{d \tau}\delta V\,,
\label{eq:ge_volume}
\end{equation}
where $\delta V$ is the infinitesimal change of volume, while for null geodesics one has
\begin{equation}
\theta \,=\,\frac{1}{\delta S}\frac{d }{d \tau}\delta S\,,
\label{eq:ge_surface}
\end{equation}
where $\delta S$ is the infinitesimal change of surface. 
 
Now we follow Ref.~\cite{Poisson:2009pwt} in order to find the Raychaudhuri's equation. We consider a set of timelike geodesics labeled by $y^a\,(a=1,2,3)$ for each point in the set. This construction therefore defines a coordinate system $(\tau,y^a)$ in a neighborhood of the geodesic $\gamma$, and there exists a transformation between this system and the one originally in use: $x^\alpha=x^\alpha(\tau,y^a)$.  We define the vectors 
\begin{equation}
e^\alpha_a\,=\, \left(\frac{\partial x^\alpha}{\partial y^a}\right)_{\tau}\,.
\end{equation}  

Given the above definitions it is legitimate to ask (as usual) that the modified Lie derivative, Eq.~\eqref{eq:lie_vec}, for the velocity vector $u^\mu=d x^\mu/d\tau$ and $e^\alpha_a$ vanishes, 
\begin{equation}
\mathcal{L}_e u^\mu\,=\,e^\nu_a u^\mu_{\,;\nu}-u^\nu e^\mu_{a\,;\nu}-\frac{\partial u^\mu}{\partial k_\alpha} \frac{\partial e^\gamma_a}{\partial x^\alpha} k_\gamma -\frac{\partial u^\mu}{\partial k_\alpha}N_{\alpha \gamma}e^\gamma_a\,=\,0\,,
\label{eq:lie_vec2}
\end{equation}
when we use the new covariant derivative of Eq.~\eqref{eq:cov_dev_st}, and we then find the following relation 
\begin{equation}
u^\nu e^\mu_{a\,;\nu}\,=\,e^\nu_a u^\mu_{\,;\nu}-\frac{\partial u^\mu}{\partial k_\alpha} \frac{\partial e^\gamma_a}{\partial x^\alpha} k_\gamma -\frac{\partial u^\mu}{\partial k_\alpha}N_{\alpha \gamma}e^\gamma_a\,.
\end{equation}

Now, it is easy to check that 
\begin{equation}
\begin{split}
\frac{d}{d \tau}\left(e^\nu_a u_\nu\right)\,=\,\left(e^\nu_a u_\nu\right)_{;\mu}u^\mu\,&=\,e^\nu_{a;\mu} u_\nu u^\mu+e^\nu_a u_{\nu;\mu}u^\mu \,=\,\left(e^\mu_a u^\nu_{\,;\mu}-\frac{\partial u^\nu}{\partial k_\alpha} \frac{\partial e^\gamma_a}{\partial x^\alpha} k_\gamma -\frac{\partial u^\nu}{\partial k_\alpha}N_{\alpha \gamma}e^\gamma_a\right) u_\nu \\
\,&=\,\frac{1}{2}\left(e^\mu_a \left(u^\nu u_\nu\right)_{\,;\mu}-\frac{\partial \left(u^\nu u_\nu\right)}{\partial k_\alpha} \frac{\partial e^\gamma_a}{\partial x^\alpha} k_\gamma -\frac{\partial\left(u^\nu u_\nu\right)}{\partial k_\alpha}N_{\alpha \gamma}e^\gamma_a\right)\,=\,0\,,
\end{split}
\end{equation}
where we have used Eqs.~\eqref{cov_der_curve}- \eqref{eq:geodesic_cov_curve} and the fact that $u^\nu u_\nu$ is constant. Then, one can always chose the parametrization of the geodesics so $e^\nu_a u_\nu=0$, being $e^\nu_a$ orthogonal to the curves. 

We now introduce a three-tensor\footnote{A three-tensor is a tensor with respect to coordinate transformations $y^a \rightarrow y^{a'}$, but a scalar with respect to transformations $x^\alpha\rightarrow x^{\alpha'}$ if $y^a$ does not change.}  $h_{ab}$ defined by 
\begin{equation}
h_{ab}\,=\, g_{\alpha\beta}e^\alpha_a e^\beta_b\,.
\end{equation}  
This acts as a metric tensor on the infinitesimal volume we are considering: for displacements confined to the cross section (so that $d\tau=0$), $x^\alpha=x^\alpha(y^a)$ and 
\begin{equation}
ds^2\,=\,g_{\alpha\beta}dx^\alpha dx^\beta\,=\,g_{\alpha\beta}\frac{\partial x^\alpha}{\partial y^a}\frac{\partial x^\beta}{\partial y^b}dy^a dy^b\,=\,h_{ab}dy^a dy^b\,.
\end{equation}  
Thus, $h_{ab}$ is the three-dimensional metric on the congruence's cross sections. Because $\gamma$ is orthogonal to its cross sections $(u_\alpha e^\alpha_a=0)$, we have that $h_{ab}=h_{\alpha\beta}e^\alpha_a e^\beta_b$ on $\gamma$, where $h_{\alpha\beta}=g_{\alpha\beta}-u_\alpha u_\beta$ is the transverse metric. If we define $h^{ab}$ to be the inverse of $h_{ab}$, then it is easy to check that 
\begin{equation}
h^{\alpha\beta}\,=\,h^{ab}e^\alpha_a e^\beta_b\,
\end{equation}  
on $\gamma$.

The three-dimensional volume element on the cross sections is $\delta V=\sqrt{h}\,d^3y$, where $h=\det (h_{ab})$. Because the coordinates $y^a$ are comoving (since each geodesic moves with a constant value of its coordinates), $d^3y$ does not change as the cross section evolves. A change in $\delta V$ therefore comes entirely from a change in $\sqrt{h}$: 
 \begin{equation}
\theta \,=\,\frac{1}{\delta V}\frac{d }{d \tau}\delta V \,=\,\frac{1}{\sqrt{h}}\frac{d }{d \tau}\sqrt{h} \,=\,\frac{1}{2}h^{ab}\frac{d h_{ab} }{d \tau}\,.
\end{equation}

We can now start computing the derivative of the three-metric:
 \begin{equation}
\theta \,=\,\frac{1}{2}h^{ab}\frac{d h_{ab} }{d \tau}\,=\,\frac{1}{2} e^a_\gamma e^b_\delta g^{\gamma\delta}g_{\alpha\beta}\left(e^\alpha_{a\,; \mu}u^\mu e^\beta_b+e^\beta_{b\,; \mu} u^\mu e^\alpha_a\right)\,=\,e^a_\alpha\left( u^\alpha_{\,;\mu}e^\mu_ a -\frac{\partial u^\alpha}{\partial k_\rho}N_{\gamma \rho}e^{\gamma}_a -\frac{\partial u^\alpha}{\partial k_\rho}\frac{\partial e^{\gamma}_a}{\partial x^\rho}k_\gamma  \right) \,=\,u^\mu_{\,;\mu}\,,
\label{eq:ray_derivation}
\end{equation}
where we have use that $e_\mu^a$ does not depend on $k$ and $e_\alpha^a  u^\alpha=0$.

So as in the GR case, we can define the tensor~\cite{Poisson:2009pwt}
\begin{equation}
B_{\alpha\beta}\,=\,u_{\alpha;\beta}\,.
\label{eq:tensor_B}
\end{equation}
Then, one can decompose the tensor $B_{\alpha\beta}$ into trace, symmetric-tracefree and antisymmetric parts
\begin{equation}
B_{\alpha\beta}\,=\,\frac{1}{3} \theta h_{\alpha\beta}+\sigma_{\alpha\beta}+\omega_{\alpha\beta}\,,
\end{equation}
where
\begin{equation}
\theta\,=\,B^\alpha _\alpha\,,\qquad\sigma_{\alpha\beta}\,=\,\frac{1}{2}\left(B_{\alpha\beta}+B_{\beta\alpha}\right)-\frac{1}{3} \theta h_{\alpha\beta}\,,\qquad \omega_{\alpha\beta}\,=\,\frac{1}{2}\left(B_{\alpha\beta}-B_{\beta\alpha}\right)\,.
\label{eq:theta1}
\end{equation}

In order to compute the Raychaudhuri's equation, we have to compute the derivative of $\theta$ with respect to $\tau$. As we showed in Eq.~\eqref{eq:geodesic_cov_curve}, the following identity holds
\begin{equation}
\frac{d \theta}{d \tau}\,=\,\theta_{;\nu}u^\nu\,=\,u^\mu_{;\mu;\nu}u^\nu\,.
\end{equation}
Now, using Eq.~\eqref{commutator_cov_der} in order to compute the commutator of two covariant derivative acting on a covariant vector, one can obtain
\begin{equation}
\frac{d \theta}{d \tau}\,=\, -B^{\alpha \beta}B_{\alpha \beta}-R_{\alpha\beta}u^\alpha u^\beta-  R_{\nu\beta\mu}(x,k)\frac{\partial u^\beta}{\partial k_\nu} u^\mu\,.
\end{equation}

From the transverse metric it is easy to check that 
\begin{equation}
u^\alpha h_{\alpha\beta}\,=\,u^\beta h_{\alpha\beta}\,=\,0\,.
\end{equation}

Using Eq.~\eqref{eq:theta1} the modified Raychaudhuri's equation for timelike geodesics can be written as
\begin{equation}
\frac{d \theta}{d \tau}\,=\,-\frac{1}{3}\theta^2-\sigma^{\mu\nu}\sigma_{\mu\nu}+\omega^{\mu\nu}\omega_{\mu\nu}-R_{\mu\nu}u^\mu u^\nu- g^{\alpha \beta} R_{\nu\beta\mu}\frac{\partial u_\alpha}{\partial k_\nu} u^\mu\,.
\label{eq:r_timelike}
\end{equation}
This last expression is the Hamiltonian version of the corresponding formula appearing in the Finslerian approach Ref.~\cite{Stavrinos:2016xyg}.

For the null case one can note that for the momentum basis we are choosing~\eqref{eq:physical_metric} a null vector for a generic curvature in spacetime is given by
\begin{equation}
u_\alpha g^{\alpha\beta}(x,k)u_\beta\,=\,0\,\implies\,u_\alpha g_x^{\alpha\beta}(x)u_\beta\,\,=\, 0\,,
\end{equation}
where, as in the previous section, $g_x^{\alpha\beta}(x)$ is the usual metric in general relativity. We see  that  in this case one can always define a null vector which does not depend on the momentum (note that this cannot be done for the timelike case) due to the fact that the original momentum metric~\eqref{eq:physical_metric} is conformally flat. For any other coordinates of the momentum metric this choice for the null vector could not be  generally  done.

Then, the procedure given in~\cite{Poisson:2009pwt} can be followed step by step due to the independence on momenta of the null vector. Introducing the usual auxiliary null vector $N_\alpha$, the transverse metric is given by~\cite{Poisson:2009pwt}
\begin{equation}
h_{\alpha\beta}\,=\, g_{\alpha\beta}-N_\alpha u_\beta-u_\alpha N_\beta\,,
\end{equation}
where
\begin{equation}
u^\beta h_{\alpha\beta}\,=\,N^\beta h_{\alpha\beta}\,=\,0\,,\qquad u^\alpha N_\alpha\,=\,1\,,\qquad N_\mu N^\mu\,=\,0.
\label{eq:n}
\end{equation}
In this case, one can define the purely traverse part of $B_{\alpha\beta}$ as   
\begin{equation}
\tilde{B}_{\alpha\beta}\,=\,h^\mu_{\alpha}h^\nu_{\beta}B_{\mu\nu}\,,
\end{equation}
whose decomposition is 
\begin{equation}
\tilde{B}_{\alpha\beta}\,=\,\frac{1}{2} \theta h_{\alpha\beta}+\sigma_{\alpha\beta}+\omega_{\alpha\beta}\,,
\end{equation}
where
\begin{equation}
\theta\,=\,\tilde{B}^\alpha _\alpha\,,\qquad\sigma_{\alpha\beta}\,=\,\frac{1}{2}\left(\tilde{B}_{\alpha\beta}+\tilde{B}_{\beta\alpha}\right)-\frac{1}{2} \theta h_{\alpha\beta}\,,\qquad \omega_{\alpha\beta}\,=\,\frac{1}{2}\left(\tilde{B}_{\alpha\beta}-\tilde{B}_{\beta\alpha}\right)\,.
\label{eq:theta2}
\end{equation}
From the commutator of two covariant derivatives one finally obtains  the  Raychaudhuri's equation for null geodesics
\begin{equation}
\frac{d \theta}{d \lambda}\,=\,-\frac{1}{2}\theta^2-\sigma^{\mu\nu}\sigma_{\mu\nu}+\omega^{\mu\nu}\omega_{\mu\nu}-R_{\mu\nu}u^\mu u^\nu\,.
\label{eq:r_null}
\end{equation}
Note that the last factor appearing in Eq.~\eqref{eq:r_timelike} here vanishes since the null vector does not depend on the momentum. So in this case the formal structure of the Raychaudhuri's equation is unchanged.

\section{Friedmann-Robertson-Walker universe}
\label{sec:universe}
In this section, we study a flat expanding universe with the momentum dependence proposed in Sec.~\ref{sec:riemann}. We start by defining our metric in the cotangent bundle with the prescription formulated in Eq.~\eqref{eq:cotangent_metric_tetrads}. We chose the space-time tetrad to be
\begin{equation}
e^0_0(x)\,=\,1\,,\qquad e^0_i(x)\,=\,e^i_0(x)\,=\,0\,,\qquad e^i_j(x)\,=\,\delta^i_j a(x^0)\,,
\label{eq:RW_tetrad_st}
\end{equation}
where $a(x^0)$ is the scale factor, and we will take the momentum de Sitter metric of Eq.~\eqref{eq:physical_metric} assuring the conservation of the Einstein tensor and validity of the second Bianchi identity. With both ingredients we are now able to construct the metric of the cotangent space
\begin{equation}
g_{00}(x,k)\,=\,\left(1-\frac{k_0^2-\vec{k}^2/a^2(x^0)}{4\Lambda^2}\right)^2\,,\qquad g_{0i}(x,k)\,=\,0\,, \qquad  g_{ij}(x,k)\,=\,\eta_{ij} a^2(x^0) \left(1-\frac{k_0^2-\vec{k}^2/a^2(x^0)}{4\Lambda^2}\right)^2\,.
\label{eq:RW_metric}
\end{equation}

One can find the Casimir from the relation~\eqref{eq:casimir_definition_cst} obtaining 
\begin{equation}
C(\bar{k})\,=\,4\Lambda^2 \arccoth^2\left(\frac{2 \Lambda}{\sqrt{k_0^2-\vec{k}^2/a^2(x^0)}}\right)\,.
\label{eq:casimir_RW}
\end{equation}
It is easy to check that when $\Lambda$ goes to infinity one recovers the usual expression of GR.

As we have mentioned in Sec.~\ref{sec:einstein}, when using a metric whose distance is proportional to $k^2$, the space-time affine connection are the same one of GR~\cite{Weinberg:1972kfs}
 \begin{equation}
H^0_{ii}(x)\,=\,a(x^0)a^\prime(x^0)\,,\qquad H^i_{0i}(x)\,=\,\frac{a^\prime(x^0)}{a(x^0)}\,,\qquad H^i_{ij}(x)\,=\,0\,,\qquad H^i_{jj}(x)\,=\,0\,,
\end{equation}
and the nonlinear coefficients are given by 
 \begin{equation}
N_{\mu\nu}(x,k)=\,k_{\lambda}H^\lambda_{\mu\nu}(x)\,.
\end{equation}

\subsection{Equations of motion from Hamilton equations}

Using the Casimir~\eqref{eq:casimir_RW} we can find the relation between the components of the momentum for the massive case (in $1+1$ dimensions for simplicity)
\begin{equation}
k_0\,=\,\frac{1}{a(x^0)}\sqrt{k_1^2 \coth\left(\frac{m}{2\Lambda}\right) + 4\Lambda^2 a^2(x^0)}\tanh\left(\frac{m}{2\Lambda}\right)\,,
\label{eq:energy_massive}
\end{equation} 
while for the massless case
\begin{equation}
k_0\,=\,\frac{k_1}{a(x^0)}\,,
\label{eq:energy_massless}
\end{equation} 
which is the same result one finds in GR. This can be understood from the fact the modification of the metric (and then of the Casimir) is proportional to the mass of the particle probing the spacetime. Then, as we will see, there will be no modification with respect to the GR computations in any calculation for massless particles. We will discuss this fact and the possible phenomenology later on.

From the first equations of~\eqref{eq:H-eqs} we can obtain the velocity to be
\begin{equation}
\dot{x}^0\,=\,\cosh^2\left(\frac{m}{2\Lambda}\right)\sqrt{\frac{k_1^2}{4 \Lambda^2 a^2(x^0)} \coth^2\left(\frac{m}{2\Lambda}\right)+1}\,,\qquad \dot{x}^1\,=\,-\frac{k_1}{2 \Lambda a^2(x^0)}\cosh^2\left(\frac{m}{2\Lambda}\right)\coth\left(\frac{m}{2\Lambda}\right)\,,
\label{eq:massive_velocities}
\end{equation} 
for the massive case, where we have used Eq.~\eqref{eq:energy_massive}, and 
\begin{equation}
 \frac{dx^0}{d\tau}\,=\,k_0\,=\,-\frac{k_1}{a(x^0)}\,,\qquad  \frac{dx^1}{d\tau}\,=\,-\frac{k_1}{ a^2(x^0)}\,,
\label{eq:massless_velocities}
\end{equation} 
for the massless case, where Eq.~\eqref{eq:energy_massless} has been applied.

Now, using the second equation of~\eqref{eq:H-eqs} we can find for massive particles that 
\begin{equation}
\ \frac{dk_0}{d\tau}\,=\,\frac{d k_0}{d x^0}\dot{x}^0 \,=\, \left(k_0^2 \coth^2\left(\frac{m}{2\Lambda}\right)-4\Lambda^2 \right)\frac{a^{\prime}(x^0)}{4 \Lambda a(x^0)}\sinh\left(\frac{m}{2\Lambda}\right)\,, \qquad   \frac{dk_1}{d\tau}\,=\,0\,.
\label{eq:momenta_RW}
\end{equation}
Solving the first equation we obtain the same expression of the energy as a function of time that we have found in Eq.~\eqref{eq:energy_massive} directly from the Casimir. 

For the massless case one arrives to 
\begin{equation}
\frac{dk_0}{d\tau}\,=\, k_0^2\frac{a^{\prime}(x^0)}{a(x^0)}\,, \qquad  \dot{k}_1\,=\,0\,.
\label{eq:momenta_RW2}
\end{equation}
This allows us to find the same relation between the energy and momentum we have found in Eq.~\eqref{eq:energy_massless}.

\subsection{Equations of motion from the metric}

In our particular case (in 1+1 for simplicity) we obtain from Eq.~\eqref{eq:horizontal_geodesics} that the geodesic equation components are
\begin{equation}
\frac{d^2x^0}{d\tau^2}+a(x^0)a^\prime(x^0)\left(\frac{dx^1}{d\tau}\right)^2\,=\,0\,,\qquad \frac{d^2x^1}{d\tau^2}+2\frac{dx^1}{d\tau}\frac{a^\prime(x^0)}{a(x^0)} \frac{dx^0}{d\tau}\,=\,0\,.
\label{eq:geo_RW}
\end{equation} 

For a massive particles, we know that the line element is 
\begin{equation}
1\,=\,\frac{dx^\mu}{d\tau}g_{\mu\nu}\frac{dx^\nu}{d\tau}\,=\,\left(1-\frac{k_0^2-\vec{k}^2/a^2(x^0)}{4\Lambda^2}\right)^2\left(\left(\frac{dx^0}{d\tau}\right)^2-a^2(x^0)\left(\frac{dx^1}{d\tau}\right)^2\right)\,.
\label{eq:velocity_RW_massive}
\end{equation} 
Then, using the relation~\eqref{eq:energy_massive} and the first equation of~\eqref{eq:geo_RW} one finds
\begin{equation}
\frac{d^2x^0}{d\tau^2}\,=\,\frac{dx^0 }{d\tau}\frac{d }{d x^0}\frac{dx^0 }{d\tau}\,=\,\frac{a'(x^0)}{a(x^0)}\cosh^4\left(\frac{m}{2\Lambda}-(u^0)^2\right)\,.
\end{equation}
Solving this differential equation one can find the same solution found in Eq.~\eqref{eq:massive_velocities} when one chooses $\tau$ to be the proper time. Another solution that we will use in the next subsection is to identify  $\tau$ as the temporal coordinate. While in GR this is possible, this is not the case in this framework since from Eq.~\eqref{eq:velocity_RW_massive} one would obtain an imaginary solution for the velocity. However, one can choose one parametrization in which $dx^0/d\tau$ does not depend explicitly on the momentum, obtaining
\begin{equation}
\frac{dx^0 }{d\tau}\,=\,\cosh^2\left(\frac{m}{2\Lambda}\right)\,.
\label{eq:massive_velocity_2}
\end{equation} 
One can see that when the high-energy scale goes to infinity one has the GR result $dx^0/d\tau=1$.

Moreover, from Eq.~\eqref{eq:horizontal_momenta}
\begin{equation}
\frac{dk_0}{d\tau}\,=\,k_1\frac{a^\prime(x^0)}{a(x^0)}\frac{dx^1}{d\tau}\,,\qquad 
\frac{dk_1}{d\tau}\,=\,k_1\frac{ a^\prime(x^0)}{a(x^0)}\frac{dx^0}{d\tau}+k_0 a(x^0)a^\prime(x^0)\frac{dx^1}{d\tau}\,.
\end{equation} 
It is easy to check that these equations lead to the same solution found in Eq.~\eqref{eq:momenta_RW}.

If we want to study photons, we have
\begin{equation}
0\,=\,\left(1-\frac{k_0^2-k_1^2/a^2(x^0)}{4\Lambda^2}\right)^2\left(\left(\frac{dx^0}{d\tau}\right)^2-a^2(x^0)\left(\frac{dx^1}{d\tau}\right)^2\right)\,,
\label{eq:velocity_RW}
\end{equation} 
and, following the same procedure of the one used for the massive case, one can find  the same results of Eq.~\eqref{eq:massless_velocities}. Also, the results obtained from Eq.~\eqref{eq:horizontal_momenta} are the same of those obtained in  Eq.~\eqref{eq:momenta_RW2} when Eq.~\eqref{eq:energy_massless} is taken into account.

We can obtain the same results starting from the Hamilton equations with a deformed Casimir, being this the squared distance in momentum space, and following the procedure proposed in Ref.~\cite{2012arXiv1203.4101M} from our definition of the nonlinear coefficients, space-time affine connection and metric in the cotangent bundle, which is consistent with what we have seen in Sec.~\ref{sec:hamilton-metric}. 

\subsection{Congruence of geodesics and Raychaudhuri's equation}
In this part we apply the results of Sec.~\ref{sec:raychaudhuri} in order to compute the congruence of geodesics and Raychaudhuri's equation for both timelike and null geodesics. 

\subsubsection*{Timelike geodesics}

Since the momentum contribution to the volume is proportional to the mass of the particle probing the spacetime, it cancels out in Eq.~\eqref{eq:ge_volume}. Then, we find the same result as in GR but with the modification term of Eq.~\eqref{eq:massive_velocity_2} 
\begin{equation}
\theta \,=\,\frac{1}{\delta V}\frac{dx^0}{d \tau}\frac{d \delta V }{d x^0}\,=\,\frac{3 a^\prime(x^0)}{2 a(x^0)} \cosh^2\left(\frac{m}{2\Lambda}\right)\,.
\label{eq:theta_massive}
\end{equation}
The choice of the $\tau$ parameter of Eq.~\eqref{eq:massive_velocity_2}  is made in order to have the same solution one finds in GR when the high-energy scale goes to infinity~\cite{Poisson:2009pwt}. 

From here we can obtain its derivative by
\begin{equation}
\frac{d \theta}{d \tau} \,=\,\frac{\delta \theta}{\delta x^\mu}u^\mu\,=\,\, 3\frac{a^{\prime\prime}(x^0)a(x^0)-a^{\prime 2}(x^0)}{a^2(x^0)} \cosh^4\left(\frac{m}{2\Lambda}\right)\,.
\label{eq:theta_massive_der}
\end{equation}

We can also compute the Raychaudhuri’s equation by considering the tangent vector field 
\begin{equation}
u_\mu\,=\,\left(1-\frac{k_0^2-\vec{k}^2/a^2(x^0)}{4\Lambda^2},0,0,0\right)\,,
\end{equation} 
so $u^\mu u_\mu=1$. From Eq.~\eqref{eq:tensor_B} and the first equation of~\eqref{eq:theta1} we obtain
\begin{equation}
\theta\,=\,\frac{12 \Lambda^2 a(x^0)a^\prime(x^0)}{\vec{k}^2-\left(k_0^2-4\Lambda^2 \right)a^2(x^0)}\,,
\end{equation}
and using Eq.~\eqref{eq:energy_massive} we find the same result of Eq.~\eqref{eq:theta_massive}.
Also, from Eq.~\eqref{eq:r_timelike} one obtains the same result of Eq.~\eqref{eq:theta_massive_der}.

We see that in these computations the only difference with respect to GR is a multiplicative factor depending on the mass of the particle.

\subsubsection*{Null geodesics}

For the null geodesics we obtain from Eq.~\eqref{eq:ge_surface}
\begin{equation}
\theta\,=\,2\frac{a^\prime(x^0)}{a^2(x^0)}\,,
\label{eq:theta_massless}
\end{equation}
and its derivative is
\begin{equation}
\frac{d \theta}{d \lambda}\,=\,2\frac{a^{\prime \prime}(x^0)a(x^0)-a^{\prime 2}(x^0)}{R^4(x^0)}\,.
\label{eq:theta_massless_der}
\end{equation}

We can now obtain the Raychaudhuri’s equation by choosing the null vector (which, as we have explained previously, cannot depend on the momentum)
\begin{equation}
u_0\,=\,\frac{1}{a(x^0)}\,\qquad u_1\,=\,1\,\qquad u_2\,=\, u_3\,=\,0\,.
\end{equation}
One can check that the geodesic equation holds
\begin{equation}
u_{\mu;\lambda}\,u^\lambda\,=\,\left(\frac{\delta u_\mu}{\delta x^\lambda}-H^\nu_{\mu\lambda}u_\nu\right)\,u^\lambda\,=\,0\,.
\end{equation}
As auxiliary null vector we can take
\begin{equation}
N_0\,=\, -N_1\,=\,\frac{\left(\vec{k}^2-\left(k_0^2-4\Lambda^2\right)a^2(x^0)\right)^2}{32 \Lambda^4 a^2(x^0)}\,\qquad N_2\,=\, N_3\,=\,0\,,
\end{equation}
in such a way that Eq.~\eqref{eq:n} holds.  As in the previous case, from Eq.~\eqref{eq:energy_massless} we find the same result of Eq.~\eqref{eq:theta_massless} and also, from Eq.~\eqref{eq:r_null} one obtains the same result of Eq.~\eqref{eq:theta_massless_der}. Summarizing, since the modification of the metrics comes basically from the mass of the particle probing the spacetime, for null geodesics there are no modifications. 

\subsection{Friedmann's equations}
We can now deal with the Einstein equations. The energy-momentum tensor for this kind of geometry is a perfect fluid
\begin{equation}
T_{\mu \nu}\,=\,u_\mu u_\nu \left(\rho(x^0)+P(x^0)\right)-P(x^0) g_{\mu \nu}(x,k)\,\qquad \text{with}\qquad u_\mu\,=\,\left(1,0,0,0\right)\,,
\label{eq:em_tensor_pf}
\end{equation}
being $u_\mu$ the fluid's four velocity and $\rho$ and $P$ the energy density and pressure of the fluid.  As we have mentioned in Sec.~\ref{sec:einstein}, for our choice of momentum metric~\eqref{eq:physical_metric} the Ricci tensor is the same one of GR and the space-time scalar of curvature can be computed from Eq.~\eqref{eq:ricci_scalar}
\begin{equation}
R\,=\,\frac{96 \Lambda^4 a^2(x^0)\left(a^{\prime 2}(x^0)+a(x^0)a^{\prime \prime}(x^0)\right)}{\left(a^2(x^0)\left(k_0 ^2-4\Lambda^2\right)-\vec{k}^2\right)^2} \,.
\end{equation}
But as we have explained in Sec.~\ref{sec:einstein}, using the momentum metric of Eq.\eqref{eq:physical_metric} there is a cancellation of momentum factors in the Einstein tensor making that this is the same one of GR. Then, the only modification of the Einstein equations are due to the modification of the energy momentum tensor. From  Eqs.~\eqref{eq:einstein_eqs} we obtain 
\begin{equation}
\begin{split}
 &a^{\prime 2}(x^0)\,=\,\frac{8 \pi}{3}\left(a^2(x^0)\left(P(x^0)+\rho(x^0)\right)+P(x^0)\frac{\left(a^2(x^0)\left(k_0 ^2-4\Lambda^2\right)-\vec{k}^2\right)^2}{16 \Lambda^4 a^2(x^0) }\right)\,,\\
&2 a^{\prime \prime}(x^0)a(x^0)+ a^{\prime 2}(x^0)\,=\,8 \pi P\frac{\left(a^2(x^0)\left(k_0 ^2-4\Lambda^2\right)-\vec{k}^2\right)^2}{16 \Lambda^4 a^2(x^0) }\,,
\end{split}
\label{eq:Friedmann_eqs}
\end{equation}

For massless particles we have the usual Friedmann equations of GR. However, this is not the case for massive particles. 

Using Eq.~\eqref{eq:casimir_RW} we can also find the relationship between the energy and the spatial components of the momentum
  \begin{equation}
k_0\,=\,\frac{\sqrt{\vec{k}^2}}{a(x^0)}\coth\left(\frac{m}{2\Lambda}\right)\tanh\left(\frac{m}{2\Lambda}\right)\,,
\end{equation}
so that Eqs.~\eqref{eq:Friedmann_eqs} becomes
\begin{equation}
\begin{split}
 &a^{\prime 2}(x^0)\,=\,\frac{8 \pi}{3}a^2(x^0)\left(P(x^0)\left(1-\sech^4 \left(\frac{m}{2\Lambda}\right)\right)+\rho(x^0)\right)\,,\\
&2 a^{\prime \prime}(x^0)a(x^0)+ a^{\prime 2}(x^0)\,=\,-8 \pi P(x^0)\sech^4 \left(\frac{m}{2\Lambda}\right)\,.
\end{split}
\label{eq:Friedmann_eqs_massive}
\end{equation}

We can now take the derivative of the first equation of~\eqref{eq:Friedmann_eqs_massive} and substitute the result in the second one, obtaining
\begin{equation}
\rho^\prime(x^0)\,=\, -3\frac{a^{\prime}(x^0)}{a(x^0)}\left(\rho(x^0)+P(x^0)\right)-\frac{a^{\prime}(x^0)}{a(x^0)}\left(1-\sech^4 \left(\frac{m}{2\Lambda}\right)\right)P^{\prime}(x^0)\,.
\end{equation}
One can check that this is the same equation for the density and pressure for the fluid that one obtains from the conservation of the energy-momentum tensor~\eqref{eq:em_tensor_pf}.

We see that the energy-momentum tensor of GR and the one we propose here~\eqref{eq:em_tensor_pf} are compatible with the Einstein equations and their conservation. From what we have studied until now there is no way to make any preference. 

However, we can use the Raychaudhuri's equation for massive particles~\eqref{eq:r_timelike} to avoid this ambiguity.  We can obtain multiplying the Einstein’s equations~\eqref{eq:einstein_tensor} by the inverse of the metric the same relationship as in GR (apart from the momentum dependence) between the scalar of curvature of spacetime and the scalar of the energy-momentum tensor
 \begin{equation}
R-2R\,=\, 8\pi T\implies\,R\,=\,- 8 \pi T\,.
\label{eq:r-t}
\end{equation}
Then, the term appearing in the Raychaudhuri's equation for massive particles~\eqref{eq:r_timelike} involving the Ricci tensor can be rewritten using Eqs.~\eqref{eq:einstein_tensor}-\eqref{eq:r-t} as 
 \begin{equation}
R_{\mu \nu} u^\mu u^\nu\,=\, \left(8 \pi T_{\mu \nu}+\frac{1}{2}g_{\mu \nu} R\right)u^\mu u^\nu\,=\,8 \pi \left( T_{\mu \nu}u^\mu u^\nu -\frac{1}{2}T\right)\,.
\end{equation}

We can now establish a relation between the congruence of geodesics and and the energy-momentum tensor
\begin{equation}
8 \pi \left( T_{\mu \nu}u^\mu u^\nu -\frac{1}{2}T\right)\,=\,\frac{d \theta}{d \tau}+\frac{1}{3}\theta^2+\sigma^{\mu\nu}\sigma_{\mu\nu}-\omega^{\mu\nu}\omega_{\mu\nu}+ g^{\alpha \beta} R_{\nu\beta\mu}\frac{\partial u_\alpha}{\partial k_\nu} u^\mu\,=\,3 \cosh^4\left(\frac{m}{2\Lambda}\right)\frac{a^{\prime\prime}(x^0)}{a(x^0)} \,.
\end{equation}
Therefore, the only way in which the previous equation can be satisfied is with a momentum dependence on the energy-momentum tensor. Indeed, using the explicit form of the energy-momentum tensor~\eqref{eq:em_tensor_pf} one obtains from the previous equation
\begin{equation}
a^{\prime\prime}(x^0)\,=\,-\frac{4\pi}{3}a(x^0)\left( P(x^0) +2 P(x^0) \sech^4\left(\frac{m}{2\Lambda}\right)+ \rho(x^0) \right)\,,
\end{equation}
which is the same expression one obtains substituting the first equation of~\eqref{eq:Friedmann_eqs_massive} into the second one. 

This is an important consistency check since, as in GR, the second Friedmann's equation can be obtained from the Einstein's equations and from the Raychaudhuri’s equation. It is obvious that for the massless case this statement also holds due to the fact that the momentum corrections vanish.

\section{Vacuum Einstein solutions}
\label{sec:vacuum}
We have seen in the previous section that for an expanding universe one can generalize the Einstein equations of GR just considering an energy-momentum tensor which depends on the momentum of the particle probing the spacetime. From our discussion of Sec.~\ref{sec:einstein}, since the Einstein tensor is not modified (it does not depend on the momenta), every vacuum solution of GR can be easily generalized in this context.  For example, the Schwarzschild solution can be easily rederived in this this framework since the Ricci and the space-time scalar of curvature vanish. 

If one wants to include a cosmological constant, one has to add to the Einstein equations~\eqref{eq:einstein_eqs} a corresponding term
 \begin{equation}
R_{\mu \nu}(x)-\frac{1}{2}g_{\mu \nu}(x,k)R(x,k)+\Lambda_c\,  g_{\mu \nu}(x,k)\,=\,\kappa T_{\mu \nu}(x,k)\,,
\label{eq:einstein_eqs_cosmological_constant}
\end{equation}
where $\Lambda_c$ is the cosmological constant one considers in GR. 

For a vacuum solution of  Einstein equations one has
 \begin{equation}
R_{\mu \nu}(x)-\frac{1}{2}g_{\mu \nu}(x,k)R(x,k)+\Lambda_c  g_{\mu \nu}(x,k)\,=\,0\,.
\label{eq:einstein_eqs_cosmological_constant_vacuum}
\end{equation}
In this case, the only difference from GR is the momentum dependence on the metric of the cosmological constant term. Then, as in we have seen in the previous section, there would be a modification in the geodesics only for massive particles.

\section{Conclusions}
\label{sec:conclusions}
In this work we have understood several formal aspects about the generalization of the well-defined description of a curved spacetime with the local group symmetry of DSR. In this scheme, we proved that one can have a consistent description of the particles motion, obtaining the same trajectories from both a deformed Hamiltonian and the geodesic motion given by a geometry in the cotangent bundle, when one identifies the Hamiltonian/Casimir as the squared distance in momentum space. From the metric formalism, we have studied the modification of the Lie derivative and shown how to include in this scheme an external force.

After that, we have defined a covariant derivative along a curve which is compatible with the geodesic motion in the same way of GR, $D u^{\mu}/D\tau=u^\mu_{;\nu}u^\nu=0$. From this definition, we obtain the geodesic deviation equation, which allows us to define a Riemann tensor in spacetime. This curvature tensor can be also obtained from the definition of the commutator of two covariant derivatives. Then, we have constructed the modified Einstein tensor and, imposing that it is conserved (in order to be conserved the energy momentum tensor), fixed a preferred system of momentum coordinates. In the literature, there is an ongoing discussion about the possibility that in DSR theories, different bases could represent different physics. This framework provides the breakdown of degeneracy of flat spacetime that appears in DSR (different possible choices of momentum basis describing the same kinematics) leading to a ``physical'' basis in which the laws of nature must be described. While this basis is obtained when a curvature on spacetime is taken into account (i.e. there is a nontrivial dependence in the space-time coordinates in the metric), in order to have a smooth limit towards GR, we argue one should consider this basis also to do phenomenology in DSR for flat spacetime.  

The obtained momentum metric is conformally flat, with a prefactor which depends on the squared four-momentum divided by the scale of deformation. This implies that the dispersion relation (the squared distance in momentum space) depends solely on the four-momentum squared, which in turn requires that the modification is carried out by a function of the squared mass divided by the high-energy scale. Hence for massless particles there would not be any modification of the trajectories while for the massive case, the deformation is suppressed by a function of the mass divided by the high-energy scale.

Using these preferred coordinates, we have computed for a modified Friedman-Robertson-Walker universe the trajectories of massive and massless particles in both Hamiltonian and geometrical cases, obtaining the same results, so corroborating our derivations presented in the first part of this work. Also, we have constructed the modified Raychaudhuri's equation in this scheme proving that, for this particular metric, one can follow the same procedure carried out in GR. Moreover, we have explicitly shown how to take into account the modification of the vacuum solutions of the Einstein equations with and without a cosmological constant.    

Finally, we can try to discuss what are the possible effects that the momentum modification in the metric provokes on the trajectory of particles and what could be the phenomenological implications. As we have seen, the massless case does not suffer any modification in the geodesic motion. However, this is not the case for the massive case. One would say that, since the modification is proportional to the mass of the particle divided by the high-energy scale, the deviation from the standard result is completely negligible. However, it is not clear in the DSR context how to consider a macroscopic system. In~\cite{Amelino-Camelia2011d} a proposal of how to consider an object made of many particles was considered. Basically, it was proposed that, if there are $N$ particles, the effective high-energy  should be $N \Lambda$. One could also think that the density of the object should play a role, making that very dense objects suffer more deviation. Then, while the modification is completely negligible for tiny particles, there could be a significant modification for really dense astrophysical objects or for a system with several components if for $N$ particles the effective scale is  $M \Lambda$ with $N \gg M$. Moreover, whereas we have understood how to consider a modified metric in the cotangent bundle for the one-particle system, we have not developed yet a proper way to take into account a multi-particle system in the geometrical scheme. This is an open issue we hope to investigate in a future work.

\section*{Acknowledgments}
This work is supported by the Spanish grants  PGC2018-095328-B-I00 (FEDER/Agencia estatal de investigación), and DGAFSE grant E21-20R. The authors would also like to thank support from the COST Action CA18108. We acknowledge useful discussions with José Manuel Carmona, José Luis Cortés and Christian Pfeifer.


\begin{thebibliography}{54}%
\makeatletter
\providecommand \@ifxundefined [1]{%
 \@ifx{#1\undefined}
}%
\providecommand \@ifnum [1]{%
 \ifnum #1\expandafter \@firstoftwo
 \else \expandafter \@secondoftwo
 \fi
}%
\providecommand \@ifx [1]{%
 \ifx #1\expandafter \@firstoftwo
 \else \expandafter \@secondoftwo
 \fi
}%
\providecommand \natexlab [1]{#1}%
\providecommand \enquote  [1]{``#1''}%
\providecommand \bibnamefont  [1]{#1}%
\providecommand \bibfnamefont [1]{#1}%
\providecommand \citenamefont [1]{#1}%
\providecommand \href@noop [0]{\@secondoftwo}%
\providecommand \href [0]{\begingroup \@sanitize@url \@href}%
\providecommand \@href[1]{\@@startlink{#1}\@@href}%
\providecommand \@@href[1]{\endgroup#1\@@endlink}%
\providecommand \@sanitize@url [0]{\catcode `\\12\catcode `\$12\catcode
  `\&12\catcode `\#12\catcode `\^12\catcode `\_12\catcode `\%12\relax}%
\providecommand \@@startlink[1]{}%
\providecommand \@@endlink[0]{}%
\providecommand \url  [0]{\begingroup\@sanitize@url \@url }%
\providecommand \@url [1]{\endgroup\@href {#1}{\urlprefix }}%
\providecommand \urlprefix  [0]{URL }%
\providecommand \Eprint [0]{\href }%
\providecommand \doibase [0]{http://dx.doi.org/}%
\providecommand \selectlanguage [0]{\@gobble}%
\providecommand \bibinfo  [0]{\@secondoftwo}%
\providecommand \bibfield  [0]{\@secondoftwo}%
\providecommand \translation [1]{[#1]}%
\providecommand \BibitemOpen [0]{}%
\providecommand \bibitemStop [0]{}%
\providecommand \bibitemNoStop [0]{.\EOS\space}%
\providecommand \EOS [0]{\spacefactor3000\relax}%
\providecommand \BibitemShut  [1]{\csname bibitem#1\endcsname}%
\let\auto@bib@innerbib\@empty
\bibitem [{\citenamefont {Feynman}(1996)}]{Feynman:1996kb}%
  \BibitemOpen
  \bibfield  {author} {\bibinfo {author} {\bibfnamefont {R.~P.}\ \bibnamefont
  {Feynman}},\ }\href@noop {} {\emph {\bibinfo {title} {{Feynman lectures on
  gravitation}}}},\ edited by\ \bibinfo {editor} {\bibfnamefont {F.~B.}\
  \bibnamefont {Morinigo}}, \bibinfo {editor} {\bibfnamefont {W.~G.}\
  \bibnamefont {Wagner}}, \ and\ \bibinfo {editor} {\bibfnamefont
  {B.}~\bibnamefont {Hatfield}}\ (\bibinfo {year} {1996})\BibitemShut {NoStop}%
\bibitem [{\citenamefont {Sahlmann}(2010)}]{Sahlmann:2010zf}%
  \BibitemOpen
  \bibfield  {author} {\bibinfo {author} {\bibfnamefont {H.}~\bibnamefont
  {Sahlmann}},\ }in\ \href
  {https://inspirehep.net/record/843661/files/arXiv:1001.4188.pdf} {\emph
  {\bibinfo {booktitle} {{Proceedings, Foundations of Space and Time:
  Reflections on Quantum Gravity: Cape Town, South Africa}}}}\ (\bibinfo {year}
  {2010})\ pp.\ \bibinfo {pages} {185--210},\ \Eprint
  {http://arxiv.org/abs/1001.4188} {arXiv:1001.4188 [gr-qc]} \BibitemShut
  {NoStop}%
\bibitem [{\citenamefont {Dupuis}\ \emph {et~al.}(2012)\citenamefont {Dupuis},
  \citenamefont {Ryan},\ and\ \citenamefont {Speziale}}]{Dupuis:2012yw}%
  \BibitemOpen
  \bibfield  {author} {\bibinfo {author} {\bibfnamefont {M.}~\bibnamefont
  {Dupuis}}, \bibinfo {author} {\bibfnamefont {J.~P.}\ \bibnamefont {Ryan}}, \
  and\ \bibinfo {author} {\bibfnamefont {S.}~\bibnamefont {Speziale}},\ }\href
  {\doibase 10.3842/SIGMA.2012.052} {\bibfield  {journal} {\bibinfo  {journal}
  {SIGMA}\ }\textbf {\bibinfo {volume} {8}},\ \bibinfo {pages} {052} (\bibinfo
  {year} {2012})},\ \Eprint {http://arxiv.org/abs/1204.5394} {arXiv:1204.5394
  [gr-qc]} \BibitemShut {NoStop}%
\bibitem [{\citenamefont {Wheeler}(1955)}]{Wheeler:1955zz}%
  \BibitemOpen
  \bibfield  {author} {\bibinfo {author} {\bibfnamefont {J.~A.}\ \bibnamefont
  {Wheeler}},\ }\href {\doibase 10.1103/PhysRev.97.511} {\bibfield  {journal}
  {\bibinfo  {journal} {Phys. Rev.}\ }\textbf {\bibinfo {volume} {97}},\
  \bibinfo {pages} {511} (\bibinfo {year} {1955})}\BibitemShut {NoStop}%
\bibitem [{\citenamefont {Rovelli}\ and\ \citenamefont
  {Speziale}(2003)}]{Rovelli:2002vp}%
  \BibitemOpen
  \bibfield  {author} {\bibinfo {author} {\bibfnamefont {C.}~\bibnamefont
  {Rovelli}}\ and\ \bibinfo {author} {\bibfnamefont {S.}~\bibnamefont
  {Speziale}},\ }\href {\doibase 10.1103/PhysRevD.67.064019} {\bibfield
  {journal} {\bibinfo  {journal} {Phys. Rev.}\ }\textbf {\bibinfo {volume}
  {D67}},\ \bibinfo {pages} {064019} (\bibinfo {year} {2003})},\ \Eprint
  {http://arxiv.org/abs/gr-qc/0205108} {arXiv:gr-qc/0205108 [gr-qc]}
  \BibitemShut {NoStop}%
\bibitem [{\citenamefont {Ng}(2011)}]{Ng:2011rn}%
  \BibitemOpen
  \bibfield  {author} {\bibinfo {author} {\bibfnamefont {Y.~J.}\ \bibnamefont
  {Ng}},\ }in\ \href
  {http://inspirehep.net/record/890213/files/arXiv:1102.4109.pdf} {\emph
  {\bibinfo {booktitle} {{Time and Matter: Proceedings, 3rd International
  Conference, TAM2010, Budva, Montenegro, 4-8 October, 2010}}}}\ (\bibinfo
  {year} {2011})\ pp.\ \bibinfo {pages} {103--122},\ \Eprint
  {http://arxiv.org/abs/1102.4109} {arXiv:1102.4109 [gr-qc]} \BibitemShut
  {NoStop}%
\bibitem [{\citenamefont {Perez}(2013)}]{Perez:2012wv}%
  \BibitemOpen
  \bibfield  {author} {\bibinfo {author} {\bibfnamefont {A.}~\bibnamefont
  {Perez}},\ }\href {\doibase 10.12942/lrr-2013-3} {\bibfield  {journal}
  {\bibinfo  {journal} {Living Rev. Rel.}\ }\textbf {\bibinfo {volume} {16}},\
  \bibinfo {pages} {3} (\bibinfo {year} {2013})},\ \Eprint
  {http://arxiv.org/abs/1205.2019} {arXiv:1205.2019 [gr-qc]} \BibitemShut
  {NoStop}%
\bibitem [{\citenamefont {Wallden}(2013)}]{Wallden:2013kka}%
  \BibitemOpen
  \bibfield  {author} {\bibinfo {author} {\bibfnamefont {P.}~\bibnamefont
  {Wallden}},\ }\bibfield  {booktitle} {\emph {\bibinfo {booktitle}
  {{Proceedings, 15th Conference on Recent Developments in Gravity (NEB 15):
  Chania, Crete, Greece, June 20-23, 2012}}},\ }\href {\doibase
  10.1088/1742-6596/453/1/012023} {\bibfield  {journal} {\bibinfo  {journal}
  {J. Phys. Conf. Ser.}\ }\textbf {\bibinfo {volume} {453}},\ \bibinfo {pages}
  {012023} (\bibinfo {year} {2013})}\BibitemShut {NoStop}%
\bibitem [{\citenamefont {Wallden}(2010)}]{Wallden:2010sh}%
  \BibitemOpen
  \bibfield  {author} {\bibinfo {author} {\bibfnamefont {P.}~\bibnamefont
  {Wallden}},\ }\bibfield  {booktitle} {\emph {\bibinfo {booktitle} {{Classical
  and quantum gravity. Proceedings, 1st Mediterranean Conference, MCCQG 2009,
  Kolymbari, Crete, Greece, September 14-18, 2009}}},\ }\href {\doibase
  10.1088/1742-6596/222/1/012053} {\bibfield  {journal} {\bibinfo  {journal}
  {J. Phys. Conf. Ser.}\ }\textbf {\bibinfo {volume} {222}},\ \bibinfo {pages}
  {012053} (\bibinfo {year} {2010})},\ \Eprint {http://arxiv.org/abs/1001.4041}
  {arXiv:1001.4041 [gr-qc]} \BibitemShut {NoStop}%
\bibitem [{\citenamefont {Henson}(2009)}]{Henson:2006kf}%
  \BibitemOpen
  \bibfield  {author} {\bibinfo {author} {\bibfnamefont {J.}~\bibnamefont
  {Henson}},\ }in\ \href@noop {} {\emph {\bibinfo {booktitle} {Approaches to
  Quantum Gravity: Toward a New Understanding of Space, Time and Matter}}},\
  \bibinfo {editor} {edited by\ \bibinfo {editor} {\bibfnamefont
  {D.}~\bibnamefont {Oriti}}}\ (\bibinfo  {publisher} {Cambridge University
  Press},\ \bibinfo {year} {2009})\ pp.\ \bibinfo {pages} {393--413},\ \Eprint
  {http://arxiv.org/abs/gr-qc/0601121} {arXiv:gr-qc/0601121 [gr-qc]}
  \BibitemShut {NoStop}%
\bibitem [{\citenamefont {Mukhi}(2011)}]{Mukhi:2011zz}%
  \BibitemOpen
  \bibfield  {author} {\bibinfo {author} {\bibfnamefont {S.}~\bibnamefont
  {Mukhi}},\ }\href {\doibase 10.1088/0264-9381/28/15/153001} {\bibfield
  {journal} {\bibinfo  {journal} {Class. Quant. Grav.}\ }\textbf {\bibinfo
  {volume} {28}},\ \bibinfo {pages} {153001} (\bibinfo {year} {2011})},\
  \Eprint {http://arxiv.org/abs/1110.2569} {arXiv:1110.2569 [physics.pop-ph]}
  \BibitemShut {NoStop}%
\bibitem [{\citenamefont {Aharony}(2000)}]{Aharony:1999ks}%
  \BibitemOpen
  \bibfield  {author} {\bibinfo {author} {\bibfnamefont {O.}~\bibnamefont
  {Aharony}},\ }\bibfield  {booktitle} {\emph {\bibinfo {booktitle} {{Strings
  '99. Proceedings, Conference, Potsdam, Germany, July 19-24, 1999}}},\ }\href
  {\doibase 10.1088/0264-9381/17/5/302} {\bibfield  {journal} {\bibinfo
  {journal} {Class. Quant. Grav.}\ }\textbf {\bibinfo {volume} {17}},\ \bibinfo
  {pages} {929} (\bibinfo {year} {2000})},\ \Eprint
  {http://arxiv.org/abs/hep-th/9911147} {arXiv:hep-th/9911147 [hep-th]}
  \BibitemShut {NoStop}%
\bibitem [{\citenamefont {Dienes}(1997)}]{Dienes:1996du}%
  \BibitemOpen
  \bibfield  {author} {\bibinfo {author} {\bibfnamefont {K.~R.}\ \bibnamefont
  {Dienes}},\ }\bibfield  {booktitle} {\emph {\bibinfo {booktitle} {{Institute
  for Theoretical Physics Conference on Unification: From the Weak Scale to the
  Planck Scale Santa Barbara, California, October 23-27, 1995}}},\ }\href
  {\doibase 10.1016/S0370-1573(97)00009-4} {\bibfield  {journal} {\bibinfo
  {journal} {Phys. Rept.}\ }\textbf {\bibinfo {volume} {287}},\ \bibinfo
  {pages} {447} (\bibinfo {year} {1997})},\ \Eprint
  {http://arxiv.org/abs/hep-th/9602045} {arXiv:hep-th/9602045 [hep-th]}
  \BibitemShut {NoStop}%
\bibitem [{\citenamefont {Philpott}\ \emph {et~al.}(2009)\citenamefont
  {Philpott}, \citenamefont {Dowker},\ and\ \citenamefont
  {Sorkin}}]{Philpott_2009}%
  \BibitemOpen
  \bibfield  {author} {\bibinfo {author} {\bibfnamefont {L.}~\bibnamefont
  {Philpott}}, \bibinfo {author} {\bibfnamefont {F.}~\bibnamefont {Dowker}}, \
  and\ \bibinfo {author} {\bibfnamefont {R.~D.}\ \bibnamefont {Sorkin}},\
  }\href {\doibase 10.1103/physrevd.79.124047} {\bibfield  {journal} {\bibinfo
  {journal} {Physical Review D}\ }\textbf {\bibinfo {volume} {79}} (\bibinfo
  {year} {2009}),\ 10.1103/physrevd.79.124047}\BibitemShut {NoStop}%
\bibitem [{\citenamefont {Liberati}(2013)}]{Liberati:2013xla}%
  \BibitemOpen
  \bibfield  {author} {\bibinfo {author} {\bibfnamefont {S.}~\bibnamefont
  {Liberati}},\ }\href {\doibase 10.1088/0264-9381/30/13/133001} {\bibfield
  {journal} {\bibinfo  {journal} {Class. Quant. Grav.}\ }\textbf {\bibinfo
  {volume} {30}},\ \bibinfo {pages} {133001} (\bibinfo {year} {2013})},\
  \Eprint {http://arxiv.org/abs/1304.5795} {arXiv:1304.5795 [gr-qc]}
  \BibitemShut {NoStop}%
\bibitem [{\citenamefont {Amelino-Camelia}(2013)}]{AmelinoCamelia:2008qg}%
  \BibitemOpen
  \bibfield  {author} {\bibinfo {author} {\bibfnamefont {G.}~\bibnamefont
  {Amelino-Camelia}},\ }\href {\doibase 10.12942/lrr-2013-5} {\bibfield
  {journal} {\bibinfo  {journal} {Living Rev.Rel.}\ }\textbf {\bibinfo {volume}
  {16}},\ \bibinfo {pages} {5} (\bibinfo {year} {2013})},\ \Eprint
  {http://arxiv.org/abs/0806.0339} {arXiv:0806.0339 [gr-qc]} \BibitemShut
  {NoStop}%
\bibitem [{\citenamefont {Belenchia}\ \emph {et~al.}(2015)\citenamefont
  {Belenchia}, \citenamefont {Benincasa},\ and\ \citenamefont
  {Liberati}}]{Belenchia:2014fda}%
  \BibitemOpen
  \bibfield  {author} {\bibinfo {author} {\bibfnamefont {A.}~\bibnamefont
  {Belenchia}}, \bibinfo {author} {\bibfnamefont {D.~M.~T.}\ \bibnamefont
  {Benincasa}}, \ and\ \bibinfo {author} {\bibfnamefont {S.}~\bibnamefont
  {Liberati}},\ }\href {\doibase 10.1007/JHEP03(2015)036} {\bibfield  {journal}
  {\bibinfo  {journal} {JHEP}\ }\textbf {\bibinfo {volume} {03}},\ \bibinfo
  {pages} {036} (\bibinfo {year} {2015})},\ \Eprint
  {http://arxiv.org/abs/1411.6513} {arXiv:1411.6513 [gr-qc]} \BibitemShut
  {NoStop}%
\bibitem [{\citenamefont {Belenchia}\ \emph {et~al.}(2016)\citenamefont
  {Belenchia}, \citenamefont {Benincasa}, \citenamefont {Liberati},
  \citenamefont {Marin}, \citenamefont {Marino},\ and\ \citenamefont
  {Ortolan}}]{Belenchia:2015ake}%
  \BibitemOpen
  \bibfield  {author} {\bibinfo {author} {\bibfnamefont {A.}~\bibnamefont
  {Belenchia}}, \bibinfo {author} {\bibfnamefont {D.~M.~T.}\ \bibnamefont
  {Benincasa}}, \bibinfo {author} {\bibfnamefont {S.}~\bibnamefont {Liberati}},
  \bibinfo {author} {\bibfnamefont {F.}~\bibnamefont {Marin}}, \bibinfo
  {author} {\bibfnamefont {F.}~\bibnamefont {Marino}}, \ and\ \bibinfo {author}
  {\bibfnamefont {A.}~\bibnamefont {Ortolan}},\ }\href {\doibase
  10.1103/PhysRevLett.116.161303} {\bibfield  {journal} {\bibinfo  {journal}
  {Phys. Rev. Lett.}\ }\textbf {\bibinfo {volume} {116}},\ \bibinfo {pages}
  {161303} (\bibinfo {year} {2016})},\ \Eprint
  {http://arxiv.org/abs/1512.02083} {arXiv:1512.02083 [gr-qc]} \BibitemShut
  {NoStop}%
\bibitem [{\citenamefont {Agullo}\ and\ \citenamefont
  {Singh}(2017)}]{Agullo:2016tjh}%
  \BibitemOpen
  \bibfield  {author} {\bibinfo {author} {\bibfnamefont {I.}~\bibnamefont
  {Agullo}}\ and\ \bibinfo {author} {\bibfnamefont {P.}~\bibnamefont {Singh}},\
  }\enquote {\bibinfo {title} {{Loop Quantum Cosmology}},}\ in\ \href {\doibase
  10.1142/9789813220003_0007} {\emph {\bibinfo {booktitle} {{Loop Quantum
  Gravity}: {The First 30 Years}}}},\ \bibinfo {editor} {edited by\ \bibinfo
  {editor} {\bibfnamefont {A.}~\bibnamefont {Ashtekar}}\ and\ \bibinfo {editor}
  {\bibfnamefont {J.}~\bibnamefont {Pullin}}}\ (\bibinfo  {publisher} {WSP},\
  \bibinfo {year} {2017})\ pp.\ \bibinfo {pages} {183--240},\ \Eprint
  {http://arxiv.org/abs/1612.01236} {arXiv:1612.01236 [gr-qc]} \BibitemShut
  {NoStop}%
\bibitem [{\citenamefont {Alesci}\ \emph {et~al.}(2017)\citenamefont {Alesci},
  \citenamefont {Botta}, \citenamefont {Cianfrani},\ and\ \citenamefont
  {Liberati}}]{Alesci:2016xqa}%
  \BibitemOpen
  \bibfield  {author} {\bibinfo {author} {\bibfnamefont {E.}~\bibnamefont
  {Alesci}}, \bibinfo {author} {\bibfnamefont {G.}~\bibnamefont {Botta}},
  \bibinfo {author} {\bibfnamefont {F.}~\bibnamefont {Cianfrani}}, \ and\
  \bibinfo {author} {\bibfnamefont {S.}~\bibnamefont {Liberati}},\ }\href
  {\doibase 10.1103/PhysRevD.96.046008} {\bibfield  {journal} {\bibinfo
  {journal} {Phys. Rev. D}\ }\textbf {\bibinfo {volume} {96}},\ \bibinfo
  {pages} {046008} (\bibinfo {year} {2017})},\ \Eprint
  {http://arxiv.org/abs/1612.07116} {arXiv:1612.07116 [gr-qc]} \BibitemShut
  {NoStop}%
\bibitem [{\citenamefont {Rovelli}\ and\ \citenamefont
  {Vidotto}(2014)}]{Rovelli:2014cta}%
  \BibitemOpen
  \bibfield  {author} {\bibinfo {author} {\bibfnamefont {C.}~\bibnamefont
  {Rovelli}}\ and\ \bibinfo {author} {\bibfnamefont {F.}~\bibnamefont
  {Vidotto}},\ }\href {\doibase 10.1142/S0218271814420267} {\bibfield
  {journal} {\bibinfo  {journal} {Int. J. Mod. Phys. D}\ }\textbf {\bibinfo
  {volume} {23}},\ \bibinfo {pages} {1442026} (\bibinfo {year} {2014})},\
  \Eprint {http://arxiv.org/abs/1401.6562} {arXiv:1401.6562 [gr-qc]}
  \BibitemShut {NoStop}%
\bibitem [{\citenamefont {Barcelo}\ \emph {et~al.}(2015)\citenamefont
  {Barcelo}, \citenamefont {Carballo-Rubio}, \citenamefont {Garay},\ and\
  \citenamefont {Jannes}}]{Barcelo:2014cla}%
  \BibitemOpen
  \bibfield  {author} {\bibinfo {author} {\bibfnamefont {C.}~\bibnamefont
  {Barcelo}}, \bibinfo {author} {\bibfnamefont {R.}~\bibnamefont
  {Carballo-Rubio}}, \bibinfo {author} {\bibfnamefont {L.~J.}\ \bibnamefont
  {Garay}}, \ and\ \bibinfo {author} {\bibfnamefont {G.}~\bibnamefont
  {Jannes}},\ }\href {\doibase 10.1088/0264-9381/32/3/035012} {\bibfield
  {journal} {\bibinfo  {journal} {Class. Quant. Grav.}\ }\textbf {\bibinfo
  {volume} {32}},\ \bibinfo {pages} {035012} (\bibinfo {year} {2015})},\
  \Eprint {http://arxiv.org/abs/1409.1501} {arXiv:1409.1501 [gr-qc]}
  \BibitemShut {NoStop}%
\bibitem [{\citenamefont {Abedi}\ \emph {et~al.}(2017)\citenamefont {Abedi},
  \citenamefont {Dykaar},\ and\ \citenamefont {Afshordi}}]{Abedi:2016hgu}%
  \BibitemOpen
  \bibfield  {author} {\bibinfo {author} {\bibfnamefont {J.}~\bibnamefont
  {Abedi}}, \bibinfo {author} {\bibfnamefont {H.}~\bibnamefont {Dykaar}}, \
  and\ \bibinfo {author} {\bibfnamefont {N.}~\bibnamefont {Afshordi}},\ }\href
  {\doibase 10.1103/PhysRevD.96.082004} {\bibfield  {journal} {\bibinfo
  {journal} {Phys. Rev. D}\ }\textbf {\bibinfo {volume} {96}},\ \bibinfo
  {pages} {082004} (\bibinfo {year} {2017})},\ \Eprint
  {http://arxiv.org/abs/1612.00266} {arXiv:1612.00266 [gr-qc]} \BibitemShut
  {NoStop}%
\bibitem [{\citenamefont {Carballo-Rubio}\ \emph {et~al.}(2018)\citenamefont
  {Carballo-Rubio}, \citenamefont {Di~Filippo}, \citenamefont {Liberati},\ and\
  \citenamefont {Visser}}]{Carballo-Rubio:2018jzw}%
  \BibitemOpen
  \bibfield  {author} {\bibinfo {author} {\bibfnamefont {R.}~\bibnamefont
  {Carballo-Rubio}}, \bibinfo {author} {\bibfnamefont {F.}~\bibnamefont
  {Di~Filippo}}, \bibinfo {author} {\bibfnamefont {S.}~\bibnamefont
  {Liberati}}, \ and\ \bibinfo {author} {\bibfnamefont {M.}~\bibnamefont
  {Visser}},\ }\href {\doibase 10.1103/PhysRevD.98.124009} {\bibfield
  {journal} {\bibinfo  {journal} {Phys. Rev. D}\ }\textbf {\bibinfo {volume}
  {98}},\ \bibinfo {pages} {124009} (\bibinfo {year} {2018})},\ \Eprint
  {http://arxiv.org/abs/1809.08238} {arXiv:1809.08238 [gr-qc]} \BibitemShut
  {NoStop}%
\bibitem [{\citenamefont {Colladay}\ and\ \citenamefont
  {Kostelecky}(1998)}]{Colladay:1998fq}%
  \BibitemOpen
  \bibfield  {author} {\bibinfo {author} {\bibfnamefont {D.}~\bibnamefont
  {Colladay}}\ and\ \bibinfo {author} {\bibfnamefont {V.~A.}\ \bibnamefont
  {Kostelecky}},\ }\href {\doibase 10.1103/PhysRevD.58.116002} {\bibfield
  {journal} {\bibinfo  {journal} {Phys. Rev.}\ }\textbf {\bibinfo {volume}
  {D58}},\ \bibinfo {pages} {116002} (\bibinfo {year} {1998})},\ \Eprint
  {http://arxiv.org/abs/hep-ph/9809521} {arXiv:hep-ph/9809521 [hep-ph]}
  \BibitemShut {NoStop}%
\bibitem [{\citenamefont {Kostelecky}\ and\ \citenamefont
  {Russell}(2011)}]{Kostelecky:2008ts}%
  \BibitemOpen
  \bibfield  {author} {\bibinfo {author} {\bibfnamefont {V.~A.}\ \bibnamefont
  {Kostelecky}}\ and\ \bibinfo {author} {\bibfnamefont {N.}~\bibnamefont
  {Russell}},\ }\href {\doibase 10.1103/RevModPhys.83.11} {\bibfield  {journal}
  {\bibinfo  {journal} {Rev. Mod. Phys.}\ }\textbf {\bibinfo {volume} {83}},\
  \bibinfo {pages} {11} (\bibinfo {year} {2011})},\ \Eprint
  {http://arxiv.org/abs/0801.0287} {arXiv:0801.0287 [hep-ph]} \BibitemShut
  {NoStop}%
\bibitem [{\citenamefont {Myers}\ and\ \citenamefont
  {Pospelov}(2003)}]{Myers:2003fd}%
  \BibitemOpen
  \bibfield  {author} {\bibinfo {author} {\bibfnamefont {R.~C.}\ \bibnamefont
  {Myers}}\ and\ \bibinfo {author} {\bibfnamefont {M.}~\bibnamefont
  {Pospelov}},\ }\href {\doibase 10.1103/PhysRevLett.90.211601} {\bibfield
  {journal} {\bibinfo  {journal} {Phys. Rev. Lett.}\ }\textbf {\bibinfo
  {volume} {90}},\ \bibinfo {pages} {211601} (\bibinfo {year} {2003})},\
  \Eprint {http://arxiv.org/abs/hep-ph/0301124} {arXiv:hep-ph/0301124}
  \BibitemShut {NoStop}%
\bibitem [{\citenamefont {Jacobson}\ and\ \citenamefont
  {Mattingly}(2001)}]{Jacobson:2000xp}%
  \BibitemOpen
  \bibfield  {author} {\bibinfo {author} {\bibfnamefont {T.}~\bibnamefont
  {Jacobson}}\ and\ \bibinfo {author} {\bibfnamefont {D.}~\bibnamefont
  {Mattingly}},\ }\href {\doibase 10.1103/PhysRevD.64.024028} {\bibfield
  {journal} {\bibinfo  {journal} {Phys. Rev. D}\ }\textbf {\bibinfo {volume}
  {64}},\ \bibinfo {pages} {024028} (\bibinfo {year} {2001})},\ \Eprint
  {http://arxiv.org/abs/gr-qc/0007031} {arXiv:gr-qc/0007031} \BibitemShut
  {NoStop}%
\bibitem [{\citenamefont {Horava}(2009)}]{Horava:2009uw}%
  \BibitemOpen
  \bibfield  {author} {\bibinfo {author} {\bibfnamefont {P.}~\bibnamefont
  {Horava}},\ }\href {\doibase 10.1103/PhysRevD.79.084008} {\bibfield
  {journal} {\bibinfo  {journal} {Phys. Rev. D}\ }\textbf {\bibinfo {volume}
  {79}},\ \bibinfo {pages} {084008} (\bibinfo {year} {2009})},\ \Eprint
  {http://arxiv.org/abs/0901.3775} {arXiv:0901.3775 [hep-th]} \BibitemShut
  {NoStop}%
\bibitem [{\citenamefont {Magueijo}\ and\ \citenamefont
  {Smolin}(2004)}]{Magueijo:2002xx}%
  \BibitemOpen
  \bibfield  {author} {\bibinfo {author} {\bibfnamefont {J.}~\bibnamefont
  {Magueijo}}\ and\ \bibinfo {author} {\bibfnamefont {L.}~\bibnamefont
  {Smolin}},\ }\href {\doibase 10.1088/0264-9381/21/7/001} {\bibfield
  {journal} {\bibinfo  {journal} {Class.Quant.Grav.}\ }\textbf {\bibinfo
  {volume} {21}},\ \bibinfo {pages} {1725} (\bibinfo {year} {2004})},\ \Eprint
  {http://arxiv.org/abs/gr-qc/0305055} {arXiv:gr-qc/0305055 [gr-qc]}
  \BibitemShut {NoStop}%
\bibitem [{\citenamefont {{Miron}}(2012)}]{2012arXiv1203.4101M}%
  \BibitemOpen
  \bibfield  {author} {\bibinfo {author} {\bibfnamefont {R.}~\bibnamefont
  {{Miron}}},\ }\href@noop {} {\bibfield  {journal} {\bibinfo  {journal} {arXiv
  e-prints}\ ,\ \bibinfo {eid} {arXiv:1203.4101}} (\bibinfo {year} {2012})},\
  \Eprint {http://arxiv.org/abs/1203.4101} {arXiv:1203.4101 [math.DG]}
  \BibitemShut {NoStop}%
\bibitem [{\citenamefont {Rund}(2012)}]{Rund2012}%
  \BibitemOpen
  \bibfield  {author} {\bibinfo {author} {\bibfnamefont {H.}~\bibnamefont
  {Rund}},\ }\href@noop {} {\emph {\bibinfo {title} {The differential geometry
  of Finsler spaces}}},\ Vol.\ \bibinfo {volume} {101}\ (\bibinfo  {publisher}
  {Springer Science \& Business Media},\ \bibinfo {year} {2012})\BibitemShut
  {NoStop}%
\bibitem [{\citenamefont {Kostelecky}(2011)}]{Kostelecky:2011qz}%
  \BibitemOpen
  \bibfield  {author} {\bibinfo {author} {\bibfnamefont {A.}~\bibnamefont
  {Kostelecky}},\ }\href {\doibase 10.1016/j.physletb.2011.05.041} {\bibfield
  {journal} {\bibinfo  {journal} {Phys. Lett.}\ }\textbf {\bibinfo {volume}
  {B701}},\ \bibinfo {pages} {137} (\bibinfo {year} {2011})},\ \Eprint
  {http://arxiv.org/abs/1104.5488} {arXiv:1104.5488 [hep-th]} \BibitemShut
  {NoStop}%
\bibitem [{\citenamefont {Stavrinos}\ and\ \citenamefont
  {Alexiou}(2017)}]{Stavrinos:2016xyg}%
  \BibitemOpen
  \bibfield  {author} {\bibinfo {author} {\bibfnamefont {P.~C.}\ \bibnamefont
  {Stavrinos}}\ and\ \bibinfo {author} {\bibfnamefont {M.}~\bibnamefont
  {Alexiou}},\ }\href {\doibase 10.1142/S0219887818500391} {\bibfield
  {journal} {\bibinfo  {journal} {Int. J. Geom. Meth. Mod. Phys.}\ }\textbf
  {\bibinfo {volume} {15}},\ \bibinfo {pages} {1850039} (\bibinfo {year}
  {2017})},\ \Eprint {http://arxiv.org/abs/1612.04554} {arXiv:1612.04554
  [gr-qc]} \BibitemShut {NoStop}%
\bibitem [{\citenamefont {Hasse}\ and\ \citenamefont
  {Perlick}(2019)}]{Hasse:2019zqi}%
  \BibitemOpen
  \bibfield  {author} {\bibinfo {author} {\bibfnamefont {W.}~\bibnamefont
  {Hasse}}\ and\ \bibinfo {author} {\bibfnamefont {V.}~\bibnamefont
  {Perlick}},\ }\href@noop {} {\  (\bibinfo {year} {2019})},\ \Eprint
  {http://arxiv.org/abs/1904.08521} {arXiv:1904.08521 [gr-qc]} \BibitemShut
  {NoStop}%
\bibitem [{\citenamefont {Girelli}\ \emph {et~al.}(2007)\citenamefont
  {Girelli}, \citenamefont {Liberati},\ and\ \citenamefont
  {Sindoni}}]{Girelli:2006fw}%
  \BibitemOpen
  \bibfield  {author} {\bibinfo {author} {\bibfnamefont {F.}~\bibnamefont
  {Girelli}}, \bibinfo {author} {\bibfnamefont {S.}~\bibnamefont {Liberati}}, \
  and\ \bibinfo {author} {\bibfnamefont {L.}~\bibnamefont {Sindoni}},\ }\href
  {\doibase 10.1103/PhysRevD.75.064015} {\bibfield  {journal} {\bibinfo
  {journal} {Phys. Rev.}\ }\textbf {\bibinfo {volume} {D75}},\ \bibinfo {pages}
  {064015} (\bibinfo {year} {2007})},\ \Eprint
  {http://arxiv.org/abs/gr-qc/0611024} {arXiv:gr-qc/0611024 [gr-qc]}
  \BibitemShut {NoStop}%
\bibitem [{\citenamefont {Amelino-Camelia}\ \emph {et~al.}(2014)\citenamefont
  {Amelino-Camelia}, \citenamefont {Barcaroli}, \citenamefont {Gubitosi},
  \citenamefont {Liberati},\ and\ \citenamefont
  {Loret}}]{Amelino-Camelia:2014rga}%
  \BibitemOpen
  \bibfield  {author} {\bibinfo {author} {\bibfnamefont {G.}~\bibnamefont
  {Amelino-Camelia}}, \bibinfo {author} {\bibfnamefont {L.}~\bibnamefont
  {Barcaroli}}, \bibinfo {author} {\bibfnamefont {G.}~\bibnamefont {Gubitosi}},
  \bibinfo {author} {\bibfnamefont {S.}~\bibnamefont {Liberati}}, \ and\
  \bibinfo {author} {\bibfnamefont {N.}~\bibnamefont {Loret}},\ }\href
  {\doibase 10.1103/PhysRevD.90.125030} {\bibfield  {journal} {\bibinfo
  {journal} {Phys. Rev.}\ }\textbf {\bibinfo {volume} {D90}},\ \bibinfo {pages}
  {125030} (\bibinfo {year} {2014})},\ \Eprint {http://arxiv.org/abs/1407.8143}
  {arXiv:1407.8143 [gr-qc]} \BibitemShut {NoStop}%
\bibitem [{\citenamefont {Lobo}\ \emph {et~al.}(2017)\citenamefont {Lobo},
  \citenamefont {Loret},\ and\ \citenamefont {Nettel}}]{Lobo:2016xzq}%
  \BibitemOpen
  \bibfield  {author} {\bibinfo {author} {\bibfnamefont {I.~P.}\ \bibnamefont
  {Lobo}}, \bibinfo {author} {\bibfnamefont {N.}~\bibnamefont {Loret}}, \ and\
  \bibinfo {author} {\bibfnamefont {F.}~\bibnamefont {Nettel}},\ }\href
  {\doibase 10.1103/PhysRevD.95.046015} {\bibfield  {journal} {\bibinfo
  {journal} {Phys. Rev.}\ }\textbf {\bibinfo {volume} {D95}},\ \bibinfo {pages}
  {046015} (\bibinfo {year} {2017})},\ \Eprint
  {http://arxiv.org/abs/1611.04995} {arXiv:1611.04995 [gr-qc]} \BibitemShut
  {NoStop}%
\bibitem [{\citenamefont {Barcaroli}\ \emph {et~al.}(2015)\citenamefont
  {Barcaroli}, \citenamefont {Brunkhorst}, \citenamefont {Gubitosi},
  \citenamefont {Loret},\ and\ \citenamefont {Pfeifer}}]{Barcaroli:2015xda}%
  \BibitemOpen
  \bibfield  {author} {\bibinfo {author} {\bibfnamefont {L.}~\bibnamefont
  {Barcaroli}}, \bibinfo {author} {\bibfnamefont {L.~K.}\ \bibnamefont
  {Brunkhorst}}, \bibinfo {author} {\bibfnamefont {G.}~\bibnamefont
  {Gubitosi}}, \bibinfo {author} {\bibfnamefont {N.}~\bibnamefont {Loret}}, \
  and\ \bibinfo {author} {\bibfnamefont {C.}~\bibnamefont {Pfeifer}},\ }\href
  {\doibase 10.1103/PhysRevD.92.084053} {\bibfield  {journal} {\bibinfo
  {journal} {Phys. Rev.}\ }\textbf {\bibinfo {volume} {D92}},\ \bibinfo {pages}
  {084053} (\bibinfo {year} {2015})},\ \Eprint
  {http://arxiv.org/abs/1507.00922} {arXiv:1507.00922 [gr-qc]} \BibitemShut
  {NoStop}%
\bibitem [{\citenamefont {Barcaroli}\ \emph {et~al.}(2017)\citenamefont
  {Barcaroli}, \citenamefont {Brunkhorst}, \citenamefont {Gubitosi},
  \citenamefont {Loret},\ and\ \citenamefont {Pfeifer}}]{Barcaroli:2016yrl}%
  \BibitemOpen
  \bibfield  {author} {\bibinfo {author} {\bibfnamefont {L.}~\bibnamefont
  {Barcaroli}}, \bibinfo {author} {\bibfnamefont {L.~K.}\ \bibnamefont
  {Brunkhorst}}, \bibinfo {author} {\bibfnamefont {G.}~\bibnamefont
  {Gubitosi}}, \bibinfo {author} {\bibfnamefont {N.}~\bibnamefont {Loret}}, \
  and\ \bibinfo {author} {\bibfnamefont {C.}~\bibnamefont {Pfeifer}},\ }\href
  {\doibase 10.1103/PhysRevD.95.024036} {\bibfield  {journal} {\bibinfo
  {journal} {Phys. Rev.}\ }\textbf {\bibinfo {volume} {D95}},\ \bibinfo {pages}
  {024036} (\bibinfo {year} {2017})},\ \Eprint
  {http://arxiv.org/abs/1612.01390} {arXiv:1612.01390 [gr-qc]} \BibitemShut
  {NoStop}%
\bibitem [{\citenamefont {Borowiec}\ and\ \citenamefont
  {Pachol}(2010)}]{Borowiec2010}%
  \BibitemOpen
  \bibfield  {author} {\bibinfo {author} {\bibfnamefont {A.}~\bibnamefont
  {Borowiec}}\ and\ \bibinfo {author} {\bibfnamefont {A.}~\bibnamefont
  {Pachol}},\ }\href {\doibase 10.1088/1751-8113/43/4/045203} {\bibfield
  {journal} {\bibinfo  {journal} {J. Phys.}\ }\textbf {\bibinfo {volume}
  {A43}},\ \bibinfo {pages} {045203} (\bibinfo {year} {2010})},\ \Eprint
  {http://arxiv.org/abs/0903.5251} {arXiv:0903.5251 [hep-th]} \BibitemShut
  {NoStop}%
\bibitem [{\citenamefont {Carmona}\ \emph {et~al.}(2019)\citenamefont
  {Carmona}, \citenamefont {Cortés},\ and\ \citenamefont
  {Relancio}}]{Carmona:2019fwf}%
  \BibitemOpen
  \bibfield  {author} {\bibinfo {author} {\bibfnamefont {J.~M.}\ \bibnamefont
  {Carmona}}, \bibinfo {author} {\bibfnamefont {J.~L.}\ \bibnamefont
  {Cortés}}, \ and\ \bibinfo {author} {\bibfnamefont {J.~J.}\ \bibnamefont
  {Relancio}},\ }\href {\doibase 10.1103/PhysRevD.100.104031} {\bibfield
  {journal} {\bibinfo  {journal} {Phys. Rev.}\ }\textbf {\bibinfo {volume}
  {D100}},\ \bibinfo {pages} {104031} (\bibinfo {year} {2019})},\ \Eprint
  {http://arxiv.org/abs/1907.12298} {arXiv:1907.12298 [hep-th]} \BibitemShut
  {NoStop}%
\bibitem [{\citenamefont {Amelino-Camelia}\ \emph
  {et~al.}(2011{\natexlab{a}})\citenamefont {Amelino-Camelia}, \citenamefont
  {Freidel}, \citenamefont {Kowalski-Glikman},\ and\ \citenamefont
  {Smolin}}]{AmelinoCamelia:2011bm}%
  \BibitemOpen
  \bibfield  {author} {\bibinfo {author} {\bibfnamefont {G.}~\bibnamefont
  {Amelino-Camelia}}, \bibinfo {author} {\bibfnamefont {L.}~\bibnamefont
  {Freidel}}, \bibinfo {author} {\bibfnamefont {J.}~\bibnamefont
  {Kowalski-Glikman}}, \ and\ \bibinfo {author} {\bibfnamefont
  {L.}~\bibnamefont {Smolin}},\ }\href {\doibase 10.1103/PhysRevD.84.084010}
  {\bibfield  {journal} {\bibinfo  {journal} {Phys. Rev.}\ }\textbf {\bibinfo
  {volume} {D84}},\ \bibinfo {pages} {084010} (\bibinfo {year}
  {2011}{\natexlab{a}})},\ \Eprint {http://arxiv.org/abs/1101.0931}
  {arXiv:1101.0931 [hep-th]} \BibitemShut {NoStop}%
\bibitem [{\citenamefont {Lobo}\ and\ \citenamefont
  {Palmisano}(2016)}]{Lobo:2016blj}%
  \BibitemOpen
  \bibfield  {author} {\bibinfo {author} {\bibfnamefont {I.~P.}\ \bibnamefont
  {Lobo}}\ and\ \bibinfo {author} {\bibfnamefont {G.}~\bibnamefont
  {Palmisano}},\ }\bibfield  {booktitle} {\emph {\bibinfo {booktitle}
  {{Proceedings, 9th Alexander Friedmann International Seminar on Gravitation
  and Cosmology and 3rd Satellite Symposium on the Casimir Effect: St.
  Petersburg, Russia, June 21-27, 2015}}},\ }\href {\doibase
  10.1142/S2010194516601265} {\bibfield  {journal} {\bibinfo  {journal} {Int.
  J. Mod. Phys. Conf. Ser.}\ }\textbf {\bibinfo {volume} {41}},\ \bibinfo
  {pages} {1660126} (\bibinfo {year} {2016})},\ \Eprint
  {http://arxiv.org/abs/1612.00326} {arXiv:1612.00326 [hep-th]} \BibitemShut
  {NoStop}%
\bibitem [{\citenamefont {Kowalski-Glikman}\ and\ \citenamefont
  {Nowak}(2002)}]{KowalskiGlikman:2002we}%
  \BibitemOpen
  \bibfield  {author} {\bibinfo {author} {\bibfnamefont {J.}~\bibnamefont
  {Kowalski-Glikman}}\ and\ \bibinfo {author} {\bibfnamefont {S.}~\bibnamefont
  {Nowak}},\ }\href {\doibase 10.1016/S0370-2693(02)02063-4} {\bibfield
  {journal} {\bibinfo  {journal} {Phys. Lett.}\ }\textbf {\bibinfo {volume}
  {B539}},\ \bibinfo {pages} {126} (\bibinfo {year} {2002})},\ \Eprint
  {http://arxiv.org/abs/hep-th/0203040} {arXiv:hep-th/0203040 [hep-th]}
  \BibitemShut {NoStop}%
\bibitem [{\citenamefont {Amelino-Camelia}(2010)}]{AmelinoCamelia:2010pd}%
  \BibitemOpen
  \bibfield  {author} {\bibinfo {author} {\bibfnamefont {G.}~\bibnamefont
  {Amelino-Camelia}},\ }\href {\doibase 10.3390/sym2010230} {\bibfield
  {journal} {\bibinfo  {journal} {Symmetry}\ }\textbf {\bibinfo {volume} {2}},\
  \bibinfo {pages} {230} (\bibinfo {year} {2010})},\ \Eprint
  {http://arxiv.org/abs/1003.3942} {arXiv:1003.3942 [gr-qc]} \BibitemShut
  {NoStop}%
\bibitem [{\citenamefont {Relancio}\ and\ \citenamefont
  {Liberati}(2020{\natexlab{a}})}]{Relancio:2020zok}%
  \BibitemOpen
  \bibfield  {author} {\bibinfo {author} {\bibfnamefont {J.~J.}\ \bibnamefont
  {Relancio}}\ and\ \bibinfo {author} {\bibfnamefont {S.}~\bibnamefont
  {Liberati}},\ }\href {\doibase 10.1103/PhysRevD.101.064062} {\bibfield
  {journal} {\bibinfo  {journal} {Phys. Rev.}\ }\textbf {\bibinfo {volume}
  {D101}},\ \bibinfo {pages} {064062} (\bibinfo {year} {2020}{\natexlab{a}})},\
  \Eprint {http://arxiv.org/abs/2002.10833} {arXiv:2002.10833 [gr-qc]}
  \BibitemShut {NoStop}%
\bibitem [{\citenamefont {Relancio}\ and\ \citenamefont
  {Liberati}(2020{\natexlab{b}})}]{Relancio:2020mpa}%
  \BibitemOpen
  \bibfield  {author} {\bibinfo {author} {\bibfnamefont {J.}~\bibnamefont
  {Relancio}}\ and\ \bibinfo {author} {\bibfnamefont {S.}~\bibnamefont
  {Liberati}},\ }\href@noop {} {\  (\bibinfo {year} {2020}{\natexlab{b}})},\
  \Eprint {http://arxiv.org/abs/2008.08317} {arXiv:2008.08317 [gr-qc]}
  \BibitemShut {NoStop}%
\bibitem [{\citenamefont {Majid}\ and\ \citenamefont
  {Ruegg}(1994)}]{Majid1994}%
  \BibitemOpen
  \bibfield  {author} {\bibinfo {author} {\bibfnamefont {S.}~\bibnamefont
  {Majid}}\ and\ \bibinfo {author} {\bibfnamefont {H.}~\bibnamefont {Ruegg}},\
  }\href {\doibase 10.1016/0370-2693(94)90699-8} {\bibfield  {journal}
  {\bibinfo  {journal} {Phys. Lett.}\ }\textbf {\bibinfo {volume} {B334}},\
  \bibinfo {pages} {348} (\bibinfo {year} {1994})},\ \Eprint
  {http://arxiv.org/abs/hep-th/9405107} {arXiv:hep-th/9405107 [hep-th]}
  \BibitemShut {NoStop}%
\bibitem [{\citenamefont {Battisti}\ and\ \citenamefont
  {Meljanac}(2010)}]{Battisti:2010sr}%
  \BibitemOpen
  \bibfield  {author} {\bibinfo {author} {\bibfnamefont {M.~V.}\ \bibnamefont
  {Battisti}}\ and\ \bibinfo {author} {\bibfnamefont {S.}~\bibnamefont
  {Meljanac}},\ }\href {\doibase 10.1103/PhysRevD.82.024028} {\bibfield
  {journal} {\bibinfo  {journal} {Phys. Rev.}\ }\textbf {\bibinfo {volume}
  {D82}},\ \bibinfo {pages} {024028} (\bibinfo {year} {2010})},\ \Eprint
  {http://arxiv.org/abs/1003.2108} {arXiv:1003.2108 [hep-th]} \BibitemShut
  {NoStop}%
\bibitem [{\citenamefont {Meljanac}\ \emph {et~al.}(2009)\citenamefont
  {Meljanac}, \citenamefont {Meljanac}, \citenamefont {Samsarov},\ and\
  \citenamefont {Stojic}}]{Meljanac:2009ej}%
  \BibitemOpen
  \bibfield  {author} {\bibinfo {author} {\bibfnamefont {S.}~\bibnamefont
  {Meljanac}}, \bibinfo {author} {\bibfnamefont {D.}~\bibnamefont {Meljanac}},
  \bibinfo {author} {\bibfnamefont {A.}~\bibnamefont {Samsarov}}, \ and\
  \bibinfo {author} {\bibfnamefont {M.}~\bibnamefont {Stojic}},\ }\href@noop {}
  {\  (\bibinfo {year} {2009})},\ \Eprint {http://arxiv.org/abs/0909.1706}
  {arXiv:0909.1706 [math-ph]} \BibitemShut {NoStop}%
\bibitem [{\citenamefont {Weinberg}(1972)}]{Weinberg:1972kfs}%
  \BibitemOpen
  \bibfield  {author} {\bibinfo {author} {\bibfnamefont {S.}~\bibnamefont
  {Weinberg}},\ }\href
  {http://www-spires.fnal.gov/spires/find/books/www?cl=QC6.W431} {\emph
  {\bibinfo {title} {{Gravitation and Cosmology}}}}\ (\bibinfo  {publisher}
  {John Wiley and Sons},\ \bibinfo {address} {New York},\ \bibinfo {year}
  {1972})\BibitemShut {NoStop}%
\bibitem [{\citenamefont {Poisson}(2009)}]{Poisson:2009pwt}%
  \BibitemOpen
  \bibfield  {author} {\bibinfo {author} {\bibfnamefont {E.}~\bibnamefont
  {Poisson}},\ }\href {\doibase 10.1017/CBO9780511606601} {\emph {\bibinfo
  {title} {{A Relativist's Toolkit: The Mathematics of Black-Hole
  Mechanics}}}}\ (\bibinfo  {publisher} {Cambridge University Press},\ \bibinfo
  {year} {2009})\BibitemShut {NoStop}%
\bibitem [{\citenamefont {Amelino-Camelia}\ \emph
  {et~al.}(2011{\natexlab{b}})\citenamefont {Amelino-Camelia}, \citenamefont
  {Freidel}, \citenamefont {Kowalski-Glikman},\ and\ \citenamefont
  {Smolin}}]{Amelino-Camelia2011d}%
  \BibitemOpen
  \bibfield  {author} {\bibinfo {author} {\bibfnamefont {G.}~\bibnamefont
  {Amelino-Camelia}}, \bibinfo {author} {\bibfnamefont {L.}~\bibnamefont
  {Freidel}}, \bibinfo {author} {\bibfnamefont {J.}~\bibnamefont
  {Kowalski-Glikman}}, \ and\ \bibinfo {author} {\bibfnamefont
  {L.}~\bibnamefont {Smolin}},\ }\href {\doibase 10.1103/PhysRevD.84.087702}
  {\bibfield  {journal} {\bibinfo  {journal} {Phys. Rev. D}\ }\textbf {\bibinfo
  {volume} {84}},\ \bibinfo {pages} {087702} (\bibinfo {year}
  {2011}{\natexlab{b}})},\ \Eprint {http://arxiv.org/abs/1104.2019}
  {arXiv:1104.2019 [hep-th]} \BibitemShut {NoStop}%
\end{thebibliography}
\end{document}